\documentclass[12pt]{article}
\usepackage[utf8]{inputenc}
\usepackage{amsmath}
\usepackage{graphicx}
\usepackage{amssymb}
\numberwithin{equation}{section}
\usepackage{mathtools}
\usepackage{siunitx} 
\usepackage{physics}
\usepackage{pdflscape}
\usepackage{diagbox}
\usepackage[usenames,dvipsnames]{color}
\usepackage{jheppub}
 \usepackage[normalem]{ulem}

\def\ZZ{{\mathbb Z}}
\def\TT{{\mathbb T}}
\def\PP{{\mathbb P}}

\def\QQ{{\mathbb Q}}
\def\OO{{\mathbb O}}
\def\RR{{\mathbb R}}
\def\CC{{\mathbb C}}

\def\Nn{{\mathcal N}}
\def\Aa{{\mathcal A}}

\date{November 2021}

\begin{document}

\baselineskip=15pt
\begin{titlepage} 
\begin{center}
{\flushright \vspace{-1.5cm}
\normalsize UUITP--56/21\\[5mm]}  
\vspace{1cm}   
\vspace*{ 2.0cm}
{\Large {\bf F-theory on 6D Symmetric Toroidal Orbifolds}}\\[12pt]
\vspace{-0.1cm}
\bigskip
\bigskip 
{ {{Finn Bjarne Kohl}$^{\,\text{a}}$}, {{Magdalena Larfors}$^{\,\text{b}}$} and 
{{Paul-Konstantin Oehlmann}$^{\,\text{b}}$}
\bigskip }\\[3pt]
\vspace{0.cm}
{\it
  ${}^{\text{a}}$ Institute for Theoretical Physics and Mathematical Institute,~Westfälische Wilhelms-Universität Münster, Wilhelm-Klemm-Straße 9, 48149 Münster, Germany\\
 ${}^{\text{b}}$ Department of Physics and Astronomy,~Uppsala University,  L\"agerhyddsv. 1, SE-751 20 Uppsala, Sweden  \\
}
\vspace{2cm}
\end{center}

\begin{abstract}
\noindent 
In this work we  study F-theory on symmetric toroidal  orbifolds that exhibit roto-translations, which are point group rotations accompanied by fractional lattice shifts.  
These geometries admit a rich class of effects, such as twisted affine folded fibers, multiple fibers, and up to three distinct torus-fibrations that yield different M/F-theory lifts.  We discuss the six-dimensional physics of the F-theory lifts, which generically host superconformal subsectors and a IIB axio-dilaton fixed to strong coupling. In addition we find that these theories exhibit a rich set of $p=0,1,2$ discrete $p$-form gauge symmetries.  
We discuss six-dimensional gauge and supergravity anomalies and match the rank and tensor branch dimension to the Hodge numbers that were computed using heterotic world sheet techniques.  
\end{abstract}

\end{titlepage}
\clearpage
\setcounter{footnote}{0}
\setcounter{tocdepth}{2}

\tableofcontents

\newpage 
\section{Introduction}
 
F-theory is a versatile framework that allows us to construct a large number of consistent string compactifications in various dimensions. Its main benefit is the possibility to go beyond perturbative IIB string constructions by geometrizing the type IIB axio-dilaton field into the complex structure modulus of an auxiliary torus \cite{Vafa_1996_1,Morrison:1996pp,Morrison:1996na}. This modulus, and thus the IIB string coupling, may then vary over the physical compactification space $B$. This provides a geometric lift of the compact space to a torus-fibration $X$, which is fixed to be Calabi-Yau in order to preserve minimal supersymmetry (SUSY). The F-theory torus on the other hand, becomes fully physical in M-theory on $X$, related to F-theory via a circle reduction.

This F-theory construction has lead to a vast progress in the construction of supergravities, superconformal field theories (SCFTs) and little string theories (LSTs) in various dimensions. In particular, F-theory has be used to classify SCFTs and LSTs in six dimensions \cite{Heckman_2015,Bhardwaj_2016} (also see references in \cite{Heckman_2019}).  Moreover, its flexibility has also lead to a systematic construction of 6D supergravity theories and the explorations of their symmetries, as described in \cite{
Kumar:2009ac,
Kumar:2009us,Park:2011ji,Morrison:2012np,Kumar:2010ru} and references therein.   

On the other hand, a particular simple set of compact geometries  are those of toroidal orbifolds. Orbifolds  are essentially flat spaces, with curvature concentrated at certain singular points. These points can often be resolved, and when this is possible, this connects the orbifold geometries to  smooth Calabi-Yau (CY) spaces. In reverse, orbifolds can often be thought of as special singular limits of smooth Calabi-Yaus in their K\"ahler moduli space. However, as there exist only a couple of hundred 6D orbifolds, in contrast to the billions of CY threefolds, this interpretation is only true for a very selected set of CY geometries. 

Due to their simplicity, toroidal orbifolds are a classic setting for string compactifications. As the flatness of the bulk of these space allows an exact heterotic world-sheet conformal field theory (CFT) description, they have been used extensively for heterotic string compactifications, starting from the first examples  \cite{DIXON1985678,DIXON1986285} and moving to large model building explorations \cite{Ibanez:1986tp,Blaszczyk:2009in,Lebedev:2006kn}.
Toroidal orbifolds have also been considered, but to a lesser extent, as backgrounds for F-theory. Orbifolds of type $(\mathbb{T}^2 \times \mathbb{T}^2 )/\ZZ_n$ for $n=2$ provided the initial example used by Sen \cite{Sen:1996vd} to relate 8D F-theory with perturbative IIB  $O7^-$ orientifold compactifications,  and this constructions was generalized to non-perturbative IIB vacua by Dasgupta and Mukhi (for $n=3,4,6$) \cite{Dasgupta_1996}. More recently, F-theory on orbifolds have played a prominent role e.g.~in the construction of 4D $\mathcal{N}=3$ SCFTs in   non-compact fourfolds \cite{Garcia-Etxebarria:2015wns,Garcia-Etxebarria:2016erx}. 

Compactifying F-theory on six dimensional toroidal orbifolds gives rise to 6D supergravity theories coupled to superconformal sectors \cite{Morrison:1996na,Gopakumar1996} 
which have recently been studied in the context of holography in \cite{hayashi2019scfts}. The latter reference considered toroidal orbifolds of type $(\mathbb{T}^2 \times \mathbb{T}^2 \times \mathbb{T}^2)/\ZZ_n \times \ZZ_m$, identifying one of the three tori as the F-theory torus. 
A key feature of F-theory compactifications is that they lead to 6D non-cricital strings obtained from D3 branes that wrap curves inside of the base $B$. For those strings a holographic IIB dual picture was proposed, that probe the $Ads_3 \times \mathbb{S}^3 \times B$ background  \cite{Haghighat:2015ega,Couzens:2017way}. Orbifold bases of the type $\mathbb{T}^4/(\ZZ_n \times \ZZ_m)$ are then again attractive and simple choices, as they are flat apart from the (codimension two) orbifold fixed points that contribute 6D tensionless strings. 
The specific constructions considered in   \cite{hayashi2019scfts} always lead to superconformal sectors that originate from the intersection of two D7 brane stacks, so called superconformal matter \cite{Heckman_2015}.

However, the studies just mentioned did not explore all possibilities. F-theory can be used to construct many more SCFT configurations and there are many more toroidal orbifolds than the examples just mentioned. In fact, 6D toroidal orbifolds have been fully classified only relatively recently \cite{Fischer_2013} leading to roughly 520 inequivalent geometries. This classification includes cases that significantly generalize those cases considered in \cite{hayashi2019scfts}, such as about 360  orbifolds with non-Abelian point group, and multiple Abelian orbifolds that admit {\it roto-translations}, i.e.~fractional lattice shifts accompanied with point group rotations. Moreover, most toroidal orbifolds admit more than one torus-fiber structure and thus allow multiple F-theory lifts that share a common M-theory dual.

The goal of this work is to initiate a systematic study of F-theory on toroidal orbifolds and determine the effective supergravity theories resulting from the compactification. Equivalently, this yields a classification of $B=\TT^2 \times \TT^2/\Gamma$ bases, relevant for their potential holographic study. In order to do so, we follow the general logic put forward in \cite{hayashi2019scfts}. In this work, we consider geometries that go beyond those considered in \cite{hayashi2019scfts} in that they can admit a single $\ZZ_n$ factor and/or roto-translational elements.  
Already with this limited generalization, we find a number of interesting geometric features that we explore   
in the F-theory setting, which translate into new ingredients in the 6D effective theory.

Toroidal orbifolds generically admit a large set of additional discrete symmetries, due to their singular nature, which can lead to a wide range of discrete R- and flavor symmetries in heterotic compactifications \cite{Nilles:2012cy}.\footnote{For a recent account for modular symmetries in orbifolds see \cite{Baur:2019kwi}.} As those discrete symmetries are inherited from the geometry, it is natural to expect enhanced {\it discrete symmetries} in F-theory physics as well. Indeed, we observe various discrete zero-, one- and two-form gauge symmetries associated to the orbifold orders. Related to that are isolated SCFT sectors in the base that are not of superconformal matter type. Moreover, we find ADE singularities that are folded by additional freely acting point group elements and yield {\it twisted affine} variants of ADE  fibers \cite{Braun:2014oya,Anderson:2019kmx}. Similarly, such twisted elements can lead to so called {\it multiple fibers} \cite{Gross97,Bhardwaj_2016,Anderson:2018heq,Anderson:2019kmx}. 
 
The relation to the heterotic string and its world-sheet CFT methods plays a crucial role in this work. In particular we use it, and its implementation into the {\it C++ Orbifolder} \cite{Nilles_2012}, to
compute stringy Hodge numbers. Those are required to determine the number of tensor and vector multiplets in the 6D supergravity theories, and to ensure anomaly cancellation.

This paper is structured as follows: In Section~\ref{sec:Orbifolds-FTheory} we give a basic introduction to F-theory from an orbifold perspective as well as a short review of orbifold geometries and their classification.  In Section~\ref{sec:results} we present our main results, summarize the F-theory features that arise from the orbifold geometry, and discuss the orbifold geometries $\ZZ_3$,  $\ZZ_3 \times \ZZ_3 -1-2$ and 
$\ZZ_3 \times \ZZ_3 -1-4$ and their physics in some detail. Some more examples, on $\ZZ_2\times\ZZ_2$ orbifolds, are discussed in Appendix \ref{sec:Fthy-Z2Z2}. In Section~\ref{sec:conclusion} we conclude and highlight future directions and open problems. 

\section{Orbifolds and F-theory}\label{sec:Orbifolds-FTheory}
\subsection{Toroidal orbifolds} 

Toroidal orbifolds play a distinctive role in string compactifications. Just like the tori they descend from, the bulk of these spaces is flat, and their curvature is concentrated at the fixed points of the orbifold action. This fact provides a lot of control to the study of e.g.~the heterotic string theory on such spaces, as the world-sheet   CFT  is free and can be solved exactly. 
Orbifolds are not isolated geometries, but can be connected, via resolution \cite{Lust:2006zh} of the orbifold singularities, to the set of smooth Calabi-Yau geometries\footnote{Such a resolution, however, might not always be possible.}. Consequently, orbifolds can be viewed as particular singular, finite-volume, limits of smooth Calabi-Yau threefolds\footnote{Oftentimes, cycles in the geometry can be shrunk further, resulting in Landau-Ginzburg, or hybrid phases of the geometry.}.  

In this section, we review the basic construction of toroidal orbifolds and how they are classified. A more detailed discussion of this matter can be found in \cite{Fischer_2013}. In the ensuing sections, we discuss F-theory and how the  orbifold geometry determines the gauge groups, field content and SCFT sectors of F-theory compactifications. 

\subsubsection{Orbifold geometry}\label{sec:orbifoldgeometry}
Toroidal orbifolds can be described as the quotient of a torus $\TT^{2n}$ with some group $G$, called the \textit{orbifolding group}, or, equivalently, as the quotient of $\RR^{2n}\simeq \CC^{n}$ with the \textit{space group} $S$ 
\begin{align} \label{eq:orbi}
	\OO_{2n} = \TT^{2n}/G  = \RR^{2n} / S = \CC^{n}/S\, .
\end{align}
   The space group is a discrete subgroup of the Poincar\'e group of $\RR^{2n}$. A generic element in $S$ is of the form $g=(\vartheta, \lambda)$ and acts on $q \in \RR^{2n}$ as
\begin{align}\label{equ:orbi-action}
	q \mapsto \vartheta q + \lambda\, ,
\end{align}
and is composed as $(\vartheta, \lambda) \circ (\omega , \tau) = (\vartheta \omega , \vartheta \tau + \lambda)$.

The subgroup of $S$ that consists of pure translations is the \textit{lattice} $\Lambda$ associated to the ${2n}$-torus $\TT^{2n} = \RR^{2n}/\Lambda$. $\Lambda$ is generated by basis vectors $e_\alpha \in \ZZ^{2n}$, $\alpha=1, ... , {2n}$. Such bases are not unique, which leads to equivalence classes of lattices. To be precise, two lattices that are invariant under the same finite unimodular group, are said to be \textit{Bravais equivalent}. Lattices can be classified in terms of their symmetry groups, and are, despite ambiguities explained in \cite{Fischer_2013}, therefore often referred to in terms of Lie algebra root-lattices. 

The elements $\vartheta$ of a space group element are discrete rotations and form the so-called \textit{point group} $P$. In case $P$ is Abelian an element $\vartheta$ is characterized by the integral vector $v\in \ZZ^{n}$ which determines the action of the rotation on the $n$ complex planes of $\CC^{n}$ according to
\begin{align}
	(\vartheta)_{ij} = \exp(2\pi \imath \frac{v_i}{r}) \; \delta_{ij}\, ,
\end{align}
where $r$ denotes the order of the point group. As described below in Section \ref{sec:OrbiClass} in order to preserve supersymmetry in string compactifications, the compact space is required to be Calabi-Yau. As a consequence, the point group needs to be a discrete subgroup of the $SU(n)$ holonomy group. This criterion translates to $\sum_i v_i/r=0\mod\ZZ$.

Note that although the space group is a symmetry of the lattice, it may still admit shifts by fractional lattice units, yielding so called roto-translations. In absence of elements with such fractional lattice shifts, the space group is given by the semi-direct product $S= P \ltimes \Lambda$. In the generic case the space group is instead described by
\begin{align}
	S = \langle\{G, \Lambda\}\rangle\, ,
\end{align}
where we recall that the orbifold group $G$ is defined in \eqref{eq:orbi}. Concluding, by slight abuse of notation, an element $g=(\vartheta, \lambda)$ in an Abelian space  group will in the following be denoted by the $(n+2n)$-component vector 
\begin{align}
	\vartheta :\; \; ((v_i/r)_{i=1, \dots , n} \, |\,  (\lambda_\alpha)_{\alpha=1, \dots , 2n})\, .
\end{align}

\begin{figure}[t]
	\centering
	\includegraphics[width=0.4\linewidth]{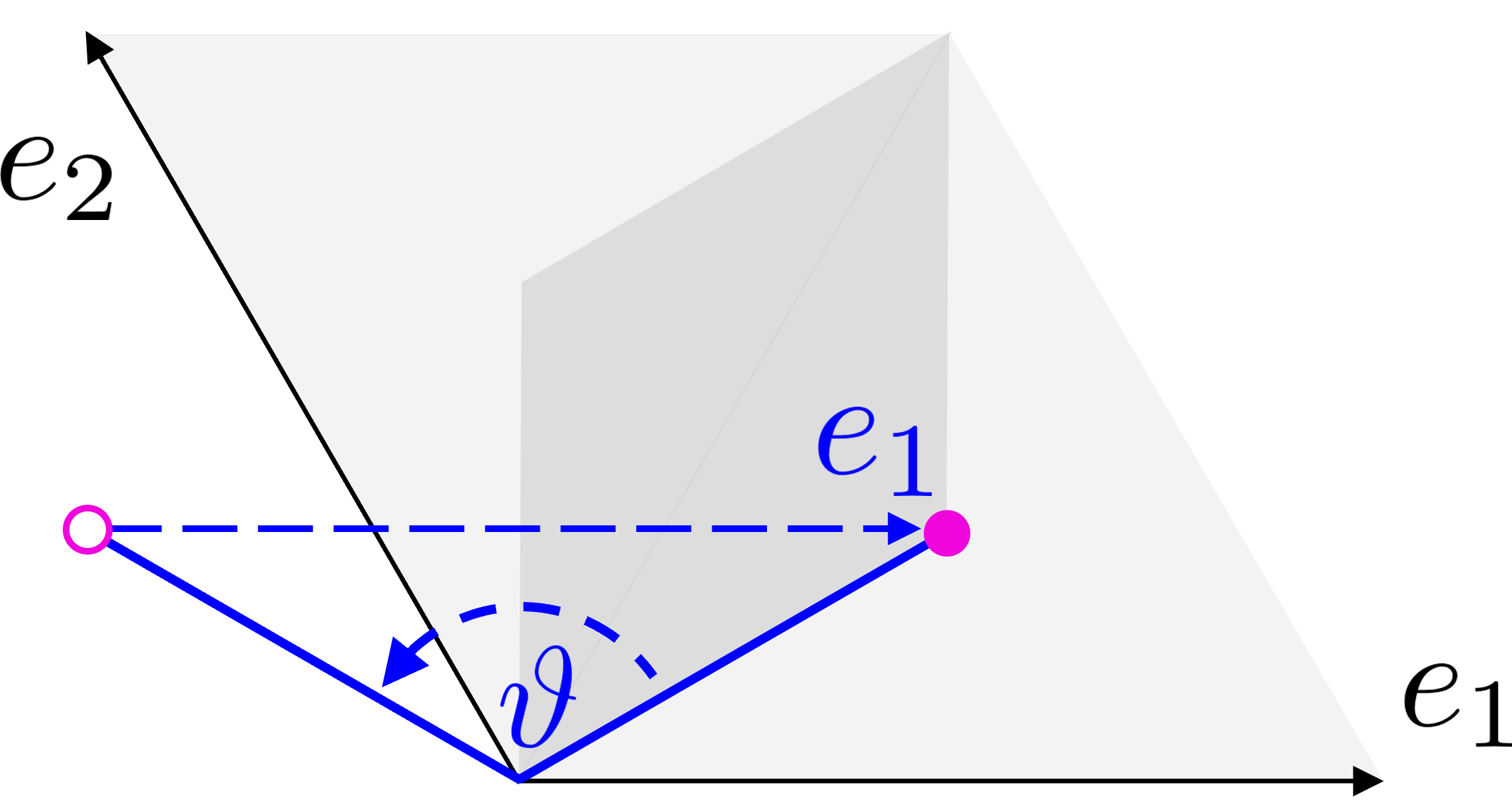}
	\caption{Depiction of the action of a pure rotation generated by $\vartheta$ by  
 $\SI{120}{\degree}$ of a an orbifold fixed point (pink) in a torus $\TT^2=\RR^2/\Lambda_{SU(3)}$. After a lattice shift by $e_1$ the point is transported back onto itself. Note, the light gray region is the fundamental domain of the lattice, opposed to the fundamental domain of the orbifold in darker gray, which is smaller due to the additional identification by the point group elements.}
	\label{fig:orbifoldaction}
\end{figure}

As indicated above, the action of the space group on $\RR^{2n}$ is in general not free and admits \textit{fixed points/lines} 
\begin{align}
\label{eq:fixedpoint}
	q_f=g(q_f)\, ,
\end{align}
as illustrated for a  $\mathbb{T}^2/\ZZ_3$ fixed point $q_f= 2/3 \, e_1 + 1/3 \,  e_2$ in Figure~\ref{fig:orbifoldaction}. Since this equation might have several solutions in the fundamental domain of the lattice for a choice of a discrete rotation, one point group element can introduce several fixed points.  Note, that only fixed points in the fundamental domain of the orbifold (cf.~Figure~\ref{fig:orbifoldaction}) are inequivalent, as we now explain.
Using \eqref{eq:fixedpoint}, one can show, that each fixed point is in one to one relation with an element $g_f$ of the orbifolding group. 
However, those fixed points $q_f$ and their corresponding group elements $g_f$ might not be independent on the full orbifold, as there might be additional group elements $h$ upon which fixed points are identified as
\begin{align}
    h (q_f) = q'_f \qquad \leftrightarrow \qquad g_f' = h\circ g_f \circ h^{-1} \, .
\end{align}     

The order $k$ of the fixed point is defined by the order of the associated point group element, such that $g^k = 1$. Furthermore, the action induces a holonomy $\Gamma \subset SU(3)$ at this point, which indicates the presence of curvature. This explains the already mentioned fact that the curvature of an orbifold is localized at the singular fixed points. Such orbifold singularities often admit a crepant resolution to a non-singular Calabi-Yau manifold, where the singular point is replaced by one, or a chain of, exceptional divisors (see e.g. \cite{Blaszczyk:2011hs,Lust:2006zh}). 
This resolved singular structure will be of crucial importance in F-theory compactifications on orbifolds, and will therefore be illustrated with two examples in Subsections~\ref{sec:geom-T4/Z3} and \ref{sec:geom-T6/Z3Z3-1-1}. Beforehand, in the next section, we review the classification of six dimensional symmetric toroidal orbifolds by Fischer and collaborators \cite{Fischer_2013}

\subsubsection{Classification of symmetric toroidal orbifolds}\label{sec:OrbiClass}
Orbifolds have been studied for a long time in the field of string theory, and several classifications exist \cite{Bailin:1999nk,Donagi:2008xy,Forste:2006wq,Dillies:2006yb}. Of particular relevance is the recent, complete classification of the symmetric toroidal orbifolds (STO)\footnote{In this work we follow the nomenclature of \cite{Fischer_2013} that distinguishes  symmetric and asymmetric orbifolds that correspond to geometric and non-geometric heterotic string backgrounds. } in six dimensions \cite{Fischer_2013}, which classifies these geometries based on their space groups. This allows one to define three types of equivalence classes among orbifold geometries
\begin{table}[t]
	\centering
\begin{tabular}{c||ccc|c}
	 $S \subset$ & $SU(3)$ & $SU(2)$ & $1$ & $\sum$ \\
	\hline\hline
	$\# \QQ$ & 17 & 4 & 1 & 22 \\
	\hline
	$\# \ZZ$ & 60 & 10 & 1 & 71 \\
	\hline
	$\# \text{aff.}$ & 138 & 23 & 1 & 162 
\end{tabular}
\caption{Summary of all orbifold geometries for all Abelian
    $\QQ$-, $\ZZ$- and affine classes split up into their respective (discrete) holonomy subgroups. Reduced holonomy subgroups lead to enhanced SUSY. }
\label{tab:orbifoldclassification} 
\end{table}
\begin{itemize}
	\item The $\QQ$\textit{-equivalence} determines the point group $P$ in the space group $S$. Two space groups are $\QQ$-equivalent iff for their point groups $P_1$ and $P_2$ 
	\begin{align}
		\exists\;  V  \in \text{GL}({2n}, \QQ): && V^{-1} P_1 V = P_2\, .&&
	\end{align}
	Orbifolds that are representatives of one $\QQ$-class have the same holonomy group. However, Bravais-inequivalent lattices are not distinguished in this classification.
	\item $\ZZ$\textit{-equivalence} distinguishes lattices $\Lambda$ in $S$. Two space groups are $\ZZ$-equivalent iff for their point groups $P_1$ and $P_2$
	\begin{align}
		\exists\;  U \in \text{GL}({2n}, \ZZ): && U^{-1} P_1 U = P_2\, .&&
	\end{align}
	The $\ZZ$-classification provides a refinement of the $\QQ$-classification, i.e.~each $\QQ$-class may contain several $\ZZ$-classes.
	\item {\it Affine-equivalence} distinguishes different orbifolding groups in the $S$, in terms of different fractional lattice shifts, so called roto-translations.  Two space groups $S_1$ and $S_2$ belong to the same affine class  
	iff
	\begin{align}
		\exists \; (f: \; \RR^{2n} \to \RR^{2n}): && f^{-1} S_1 f = S_2\, ,&&
	\end{align}
	with $f = (A, t)$ consisting of translations $t$ and linear mappings $A$.\\ 
	This equivalence is of special importance in this work as the effect of different \textit{non-trivial} affine classes on the physical theory is investigated.
\end{itemize}
\noindent
In Ref.~\cite{Fischer_2013}, these equivalence relations are used to classify and count inequivalent  orbifolds, grouped in terms of the amount of supersymmetry preserved in compactification of the heterotic string to four dimensions. We summarize this classification, for Abelian point groups,   in Table~\ref{tab:orbifoldclassification}. For a more detailed analysis of the classification the reader is referred to the original work \cite{Fischer_2013}.

In string compactifications, the geometry of the internal space determines the theory in lower dimensions, and so these different equivalence classes determine different aspects of the physical theory. Whereas the $\QQ$-class determines the amount of SUSY that is preserved in the compactified theory, transforming between different representatives of one affine class corresponds to moving in the moduli space of the respective compactification.

Note, that in the classification as well as in the present work geometries of both simply-connected and non-simply connected geometries appear. The fundamental group of toroidal orbifolds is given by 
\begin{align}\label{equ:fundgroup}
    \pi_1 = S/\expval{F}\, ,
\end{align}
where $\expval{F}$ denotes the span of non-freely acting space group elements. In case $\expval{F}=S$ (which holds for most affine classes), $\pi_1$ is trivial. A non-trivial fundamental group exists, if there are freely acting space group elements, that are not decomposable into the set of non-freely acting elements. In the context of the heterotic string this indicates non-local symmetry breaking. In the present work the orbifold discussed in Section~\ref{subsec:z3z3-1-4} is an example for an orbifold with non-trivial fundamental group. As we will see later, similarly as discussed in \cite{Anderson:2018heq,Anderson:2019kmx} such geometries lead to F-theory description in terms of a covering theory modulo   additional superconformal sectors. 

In the following, adopting the nomenclature of the classification, six dimensional STOs are referred to in terms of these three classes as $\ZZ_n\times \ZZ_m -X-Y$, where $\ZZ_n\times \ZZ_m$ specifies the point group, that is the $\QQ$-class, and $X$ and $Y$ determine the $\ZZ$- and affine class, respectively. 

\subsubsection{Stringy Hodge numbers of orbifolds}
\label{sec-hodge-orbifold} 
A key component in the analysis of F-theory compactifications are the K\"ahler and complex structure moduli of the orbifold geometries $X_{\text{orb}}$. Strictly speaking, these topological invariants are not defined for singular geometries but only for the smooth Calabi-Yau 
A short-cut to determining these numbers, that bypasses the resolution of the geometry, is provided by the free heterotic CFT in the standard embedding. This refers to a particular solution of the Bianchi identity and Hermitian Yang-Mills equation of the heterotic compactification, where the gauge connection is equated with the 
  $SU(3)$ spin connection. The observable 4D gauge group is then the $SU(3)$ commutant in $E_8 \times E_8$, that is $E_6 \times E_8$, and we can use this information to determine the Hodge numbers. Decomposing the broken $E_8$ adjoint as
\begin{align}
	\mathbf{248} \to  (\mathbf{1}, \mathbf{8}) + ( \mathbf{27}, \mathbf{3}) + (\overline{\mathbf{27}}, \overline{\mathbf{3}}) + (\mathbf{78}, \mathbf{1})\, .
\end{align}
we observe that $\mathbf{27}$ and $\mathbf{\overline{27}}$ come with a $\mathbf{3}$ and $\bar{\mathbf{3}}$ representations of the $SU(3)$ tangent bundle $T_X$ of the CY $X$. Hence, one can count the $\mathbf{27}$, $\mathbf{\overline{27}}$ representations\footnote{Recall that the $\overline{\mathbf{3}}$ is the double antisymmetric of the fundamental of $SU(3)$.} as 
\begin{align}
\mathbf{27}: \text{dim}(H^1(X,T_X)) = h^{(1,1)}(X) \, , \quad \overline{\mathbf{27}}: \text{dim}(H^1(X,\wedge^2 T_X)) = h^{(1,2)}(X) \, ,
\end{align}
which are thus the Hodge number of the threefold. These target space fields also correspond to the (a,c) and (c,c) chiral primaries of the heterotic $(2,2)$ world sheet theory. Such operators are preserved in different phases of the target space geometry, as explicitly given via the GLSM construction \cite{Witten:1993yc}. 

Hence if we can compute the (anti-)fundamentals of $E_6$ of the heterotic target space theory in the standard embedding, we know the Hodge numbers also in the smooth phase. Luckily, we have such a method as the heterotic world sheet theory is a free CFT at the orbifold point and the spectrum can thus be computed exactly. This has conveniently been implemented in the  {\it C++ Orbifolder} \cite{Nilles_2012}.
This information also allows to deduce the amount of twisted and untwisted moduli in terms of string states that are, respectively, located at the fixed points and freely propagating bulk states. These states correspond to the deformations of the fixed points and the parent  tori respectively. 
\begin{figure}[t]
	\centering
	\includegraphics[width=0.4\linewidth]{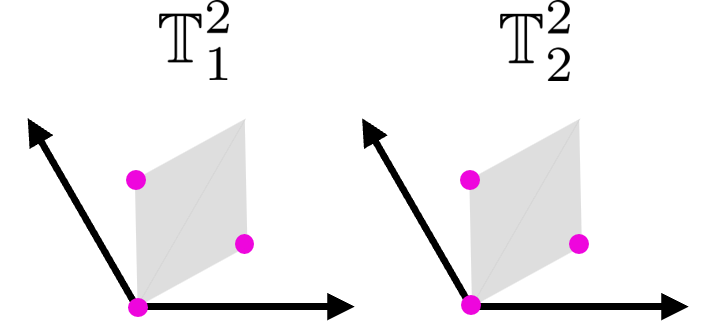}
	\caption{This depicts the location of the fixed points (pink) of the orbifold $(\TT^2\times\TT^2)/\ZZ_3$. The lattice $\Lambda=SU(3)\times SU(3)$ is depicted in terms of its generating vectors (black) and the fundamental domain of the orbifold is marked gray.}
	\label{fig:t4-z3-fpintori}
\end{figure}
\noindent
For all the aforementioned Abelian orbifolds, this information is summarized in Appendix C.1 of \cite{Fischer_2013}.

We exemplify those considerations with a two- and three-fold example.

\subsubsection*{Geometry of $\TT^4/\ZZ_3$:} \label{sec:geom-T4/Z3}
Orbifolds of type $\TT^{2n}/\ZZ_3$ have been extensively studied in the string theory literature since \cite{DIXON1985678}. As all four dimensional non-trivial Calabi-Yau orbifolds are K3 surfaces,
this example is a singular limit of an (elliptic) K3 as e.g. considered by Sen in \cite{Sen:1996vd} and \cite{Dasgupta_1996}. 
  
\begin{figure}[t]
	\centering
	\includegraphics[width=0.6\linewidth]{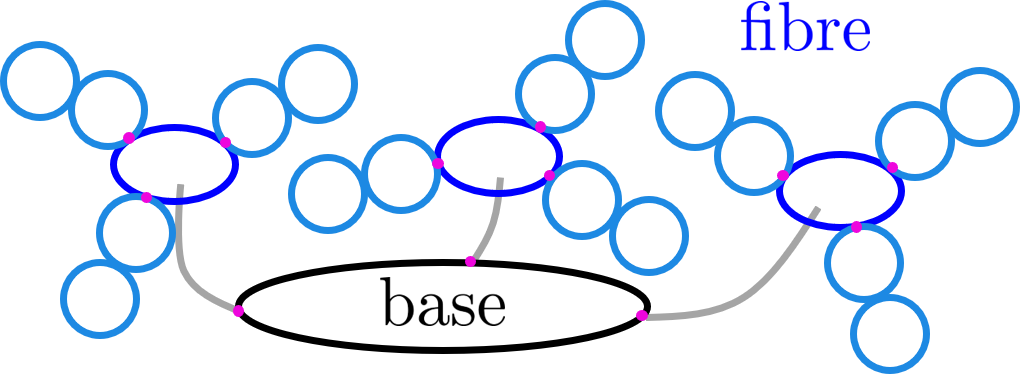}
	\caption{Depiction of the resolution of the orbifold singularities in $(\TT^2\times\TT^2)/\ZZ_3$. Light blue circles depict the $\ZZ_3$ resolution divisors. The ambient tori over fiber and base also become $\mathbb{P}^1$'s depicted as blue and black circles. }
	\label{fig:t4-z3-fpfibration}
\end{figure}

To be explicit, this geometry has one point group generator of the form
\begin{align}
	\begin{array}{cccc|ccccc}
		\vartheta: &&  \left(\right.1/3 & -1/3 \;\; &\; 0 & 0 & 0 & 0 \left.\right) 
	\end{array}\, .
\end{align}
This acts on a lattice $\Lambda=SU(3) \times SU(3)$ which is the only lattice invariant under a $\ZZ_3$ action. The action of this element induces three fixed locations in each of the two tori giving rise to in total nine fixed points. This structure is depicted in Figure~\ref{fig:t4-z3-fpintori}. The order of these fixed points is three, requiring two blow-ups at each fixed point. 
The resolved geometry is depicted in Figure~\ref{fig:t4-z3-fpfibration} where the black circle denotes the base $\mathbb{P}^1$, the dark blue circle the generic F-theory fiber and the light blue circles the blown up singularities. There we find over each of the three base loci an $E_6$ type fiber which required $3\times 6=18$ resolution divisors. This geometry is in fact an extremal K3 surface, which can be seen as follows: The Picard number $\rho$ counts the subpart of $h^{(1,1)}(K3)$ that is realized as divisors modulo algebraic equivalence. Extremal K3s are rigid, i.e. they have no other deformations but the K\"ahler moduli associated to divisors. Since $h^{(1,1)}(K3)=20$, it then follows that $\rho=20$.  This is exactly what we have in our case since  $\rho=2+18=20$, where the $2$ is the contribution of fiber and base.  

We can double check this observation from the (4,4) heterotic orbifold CFT perspective as analyzed in \cite{Honecker:2006qz}. The standard embedding
requires us to identify the $SU(2)$ gauge group inside one $E_8$ with the spin connection. This 
breaks the heterotic six dimensional target space gauge group as $E_8 \rightarrow E_7$. Each $ \mathbf{56}$ half-hypermultiplet in 6D  represents a deformation in the Picard group of the K3. For this geometry there are two half-hypers in the bulk, and two $\mathbf{56}$ at each of the nine fixed points
\cite{Honecker:2006qz}. In total we have the twisted and untwisted Picard numbers
\begin{align}
    (\rho_{\text{untw}},\rho_{\text{tw}}) = ( 2, 18) \, , 
\end{align} 
which shows that we have correctly identified the singularities.
A notable benefit of extremal elliptic K3s is, that they are classified in \cite{shimada2001} which we will also use in order to discuss the F-theory physics in Section~\ref{ssec:K3z3Ftheory}.  

\subsubsection*{Geometry of $\TT^6/\ZZ_3\times\ZZ_3-1-1$} \label{sec:geom-T6/Z3Z3-1-1}
Turning to six dimensional STOs, there exists the option to have not only one point group factor, but two (and accepting non-Abelian orbifolds also three). As an example, take the orbifold of $\TT^6=\RR^6/\Lambda_{SU(3)^3}$ modulo two $\ZZ_3$ quotients. The generators $\vartheta$ and $\omega$ of the two quotients can be chosen to act according to 
\begin{align}
\label{eq:Z3Z311}
	\begin{array}{ccccc|ccccccc}
		\vartheta: && \left(\right.\;\;0\;\;\; & 1/3 & -1/3 \;\;&\; 0 & 0 & 0 & 0 & 0 & 0 \left.\right) & \\
		\omega: && \left(\right.1/3 & 0 & -1/3 \;\;&\; 0 & 0 & 0 & 0 & 0 & 0 \left.\right) & .
	\end{array}
\end{align}
\begin{figure}[t]
	\centering
	\includegraphics[width=0.6\linewidth]{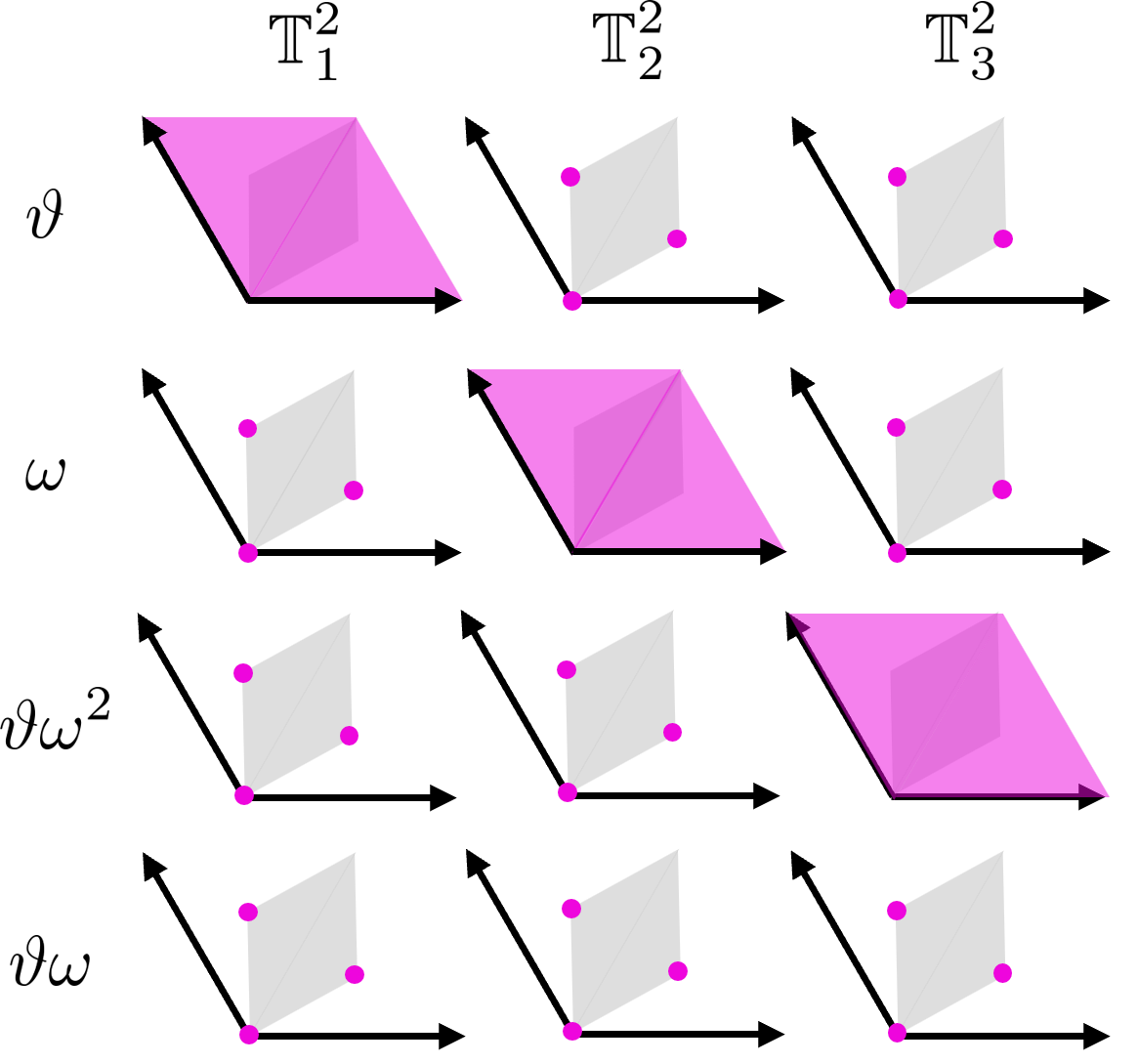}
	\caption{This depicts the location of the fixed points (pink) of the orbifold $(\TT^2\times\TT^2\times\TT^2)/\ZZ_3\times\ZZ_3-1-1$ grouped in sectors according to the group element that fixes them. The lattice $\Lambda=SU(3)\times SU(3)\times SU(3)$ is depicted in terms of its generating vectors (black) and the fundamental domain of the orbifold is marked gray.}
	\label{fig:z3xz3fpintori}
\end{figure} 

The generated inequivalent group elements induce fixed points, that are depicted in Figure~\ref{fig:z3xz3fpintori}. The fixed points may be grouped according to the element that fixes them. There are 27 fixed points associated to the group element $\vartheta\omega$ that are localized in all three tori (i.e.~they are co-dimension three). In contrast to these, the remaining inequivalent group elements all have fixed planes, that is its action leaves an entire torus fixed. The fixed points associated to these group elements are referred to as co-dimension two. Analyzing these sectors separately, one realizes that each of them resembles the orbifold $\TT^4/\ZZ_3 \times \TT^2$, where the first factor is the four dimensional orbifold discussed in the previous chapter. The second factor is either the first, second or third torus of $\TT^6=\TT^2\times\TT^2\times\TT^2$ for each of the three group elements. Note, that the fixed points in these three sectors mutually intersect in the co-dimension three fixed points associated to $\vartheta\omega$ discussed above. 

Regarding the Hodge numbers, one finds no contributions to $h^{(2, 1)}$. Furthermore, there are three untwisted states contributing to $h^{(1, 1)}$. The $27$ fixed points in the sector associated to $\vartheta\omega$ of co-dimension three contribute each one state twisted state to $h^{(1, 1)}$. Due to the states associated to the remaining sectors one finds \cite{Fischer_2013}
\begin{align}
	(h^{(1, 1)}_{\text{untw.}}, h^{(1, 1)}_{\text{tw.}}) = (3, 78)\, , && 	(h^{(2, 1)}_{\text{untw.}}, h^{(2, 1)}_{\text{tw.}}) = (0, 0)\, .
\end{align}

\subsection{F-theory and SCFTs}\label{sec:Ftheory}

In this section we summarize the basic concepts of F-theory, and how this framework allows to construct 6D superconformal field theories (SCFTs). A more thorough summary can be found in one of the many reviews in the literature (see e.g.\  \cite{weigand2018tasi,Heckman_2019}). F-theory means that we enhance the IIB compactification space $B_n$ by an auxiliary torus, which we refer to as the F-theory torus in the following, whose complex structure describes the IIB axio-dilaton coupling $\tau = C_0 + i g_{\text{IIB}}$ modulo $SL(2,\mathbb{Z})$, which is a symmetry of the type IIB theory.  This geometrization  allows $\tau$ to vary from point  to point over the compactification space, as described by a holomorphic fibration.  The total space $X_{n+1}$ hence obtains the structure
\begin{align}
    \begin{array}{cc}
    \TT^2 \rightarrow& X_{n+1} \\
    &\downarrow \pi \\
    & B_n
    \end{array} \, .
\end{align}
The power of F-theory thus lies in the possibility tp
olet the axio-dilaton vary, which allows to explore  strong coupling regimes.

When compactified on a circle, F-theory is dual to M-theory on $X_{n+1}$, where the volume of the elliptic fiber becomes physical. In order to preserve minimal SUSY of the M-theory compactification, $X_{n+1}$ is fixed to be Calabi-Yau. This perspective gives a dual {\it definition} of F-theory, by thinking of it as M-theory on the CY $(n+1)$-fold upon decompactifying the elliptic fiber. This process we call the {\it F-theory lift} with respect to a specified torus fiber of a given $(n+1)$-fold.  
The power of the construction is, that we are able to keep track and read off the IIB axio-dilaton $\tau$ in a SUSY preserving way, for cases when it is neither constant nor fixed to small couplings. This is important when adding D7 branes, that act as defects that source $SL(2,\ZZ)$ valued monodromies for the axio-dilaton $\tau$. While perturbative D7 branes shift $\tau$ by the $T$  generator of $SL(2,\ZZ)$, this generalization also allows for strong coupling branes that shift $\tau$ by the $S$ generator.

When moving onto the locus of a D7 brane, the axio-dilaton $\tau$ diverges and in turn the F-theory $\TT^2$ becomes singular. These singularities admit an ADE classification which, in type IIB language, describes the D7 worldvolume gauge algebra $G$. This connection is beautifully reproduced in geometry, where the intersection matrix of the resolved fiber is precisely the (affine) ADE Cartan matrix. In 8D, such singularities are schematically of form
\begin{align}
\label{eq:Y2}
    Y_{2} = (\mathbb{T}^2 \times \mathbb{C})/ \Gamma_n \, ,\qquad \Gamma_n \subset SU(2)\, \quad \text{ with } (\Gamma_n)^n=1 \, .
\end{align}
 E.g. for $\Gamma_3 = (e^{2 \pi i/3}, e^{-2 \pi i/3})$ this yields three order three fixed points in the $\TT^2$ direction and the origin of the $\mathbb{C}$ factor. Each are resolved by two $(-2) $ curves in $\mathbb{T}^2$, leading to an affine $E_6$ over the origin of $\mathbb{C}$.
  
 This geometric correspondence guarantees the very same gauge algebra to appear
 in M-theory, when compactified on the same (singular) geometry, as it is dual to F-theory reduced on a circle. In M-theory the torus-fiber and its resolution is physical and the volume of the resolution curves are given by  K\"ahler moduli. Expanding the M-theory 3-form along those $(1,1)$-forms associated to K\"ahler moduli, contributes vector multiplets that yield the $U(1)^{\text{rk}(G)}$ Cartan algebra. At the singular point this Cartan subalgebra is enhanced to the full non-Abelian group, that originates from M2 branes that wrap the collapsing fibral divisors.

\subsubsection{6D anomaly cancellation}
\label{sec:anomalies} 
Any consistent quantum theory must be free of gauge and gravity anomalies, which can be achieved via the GSSW mechanism  \cite{Green:1984bx,Sagnotti:1992qw}. In the following we are mostly interested in 6D $\mathcal{N}=(1,0)$ supergravity theories, where the anomaly cancellation conditions are particularly tight. In the following, we briefly review those constraints, and refer to \cite{taylor2011tasi} or \cite{Kumar:2010ru,Johnson_2016} for more details. We will then discuss how these conditions are encoded in the geometry of the compactification.

Any supergravity theory must be free of the irreducible gravitational anomaly, which translates to the condition
\begin{align}
\label{eq:gravanomaly}
     H-V+29T=273 \, ,
\end{align}
where $H,V$ and $T$ are the numbers of 6D massless hyper, vector and tensor multiplets, respectively. Then there are the reducible gauge and gravitational anomaly cancellation conditions  
\begin{align}
    a \cdot a &= 9-T \\
    a \cdot b &= - \frac{\lambda}{6} \left( A_{adj}-\sum_{\mathbf{R}} n[\mathbf{R}] A_\mathbf{R}  \right)\\
    b \cdot b &= - \frac{\lambda^2}{3} \left( C_{adj}-\sum_{\mathbf{R}} n[\mathbf{R}] C_\mathbf{R}   \right)\\
    0&= B_{adj}-\sum_{\mathbf{R}} n[\mathbf{R}] B_\mathbf{R}  \; ,
\end{align}
where the right hand side are the anomalies, which should be cancelled by a GS contribution on the left.\footnote{Additional sets of conditions arise when $U(1)$ gauge factors are present. We leave these out for brevity, and refer the interested reader to e.g. Appendix~A of \cite{Klevers:2014bqa}.}
On the right hand side, $n[\mathbf{R}]$ is the number of hypermultiplets in a representation $\mathbf{R}$ in the theory, and $A_\mathbf{R},B_\mathbf{R},C_\mathbf{R}$ are Casimir coefficients of representations $\mathbf{R}$ of gauge algebras\footnote{
In this work we do not distinguish gauge groups and algebras. If those structures do no coincide, we remark so explicitly.
} $G$. On the left hand side, the coefficients $a$ and $b$ are $SO(1,T)$  vectors, and $\cdot$ denotes the scalar product with the $SO(1,T)$ invariant pairing $\Omega_{ij}$. This tensor determines the GS axion couplings as well as the coupling of BPS strings to the tensor multiplets in the theory with the condition to be uni-modular \cite{Seiberg:2011dr}, that is $\text{Det}(\Omega)=-1$.

All  field theory quantities above can be matched to the geometric quantities of F-theory on $X_3$. In particular the pairing $\Omega$  can be expressed as the intersection pairing on the base $B_2$
\begin{align}
    \Omega_{ij} = \omega_i \cdot \omega_j
\end{align}
with  $\omega \in H^2(B_2, \mathbb{Z})$ being an integral basis of 2-cycles. Further, $b$ and $a$ correspond to divisors $\mathcal{Z}$ of the codimension one singularities and the anticanonical class $c_1(B_2)$ of the base respectively. The $h^{(1,1)}(B_2)=1+T$ dimensional base cohomology lifts to the vertical divisor part of the the threefold $X_3$. The Shioda-Tate-Wazir formula tells us then, that all other K\"ahler moduli of the threefold $X_3$ are of fibral type or sections. Thus, they count ADE resolution divisors and  generators of a free Mordell-Weil (MW) group, respectively. It is by now well established that the former class lifts to  ADE  gauge algebras and the second to $U(1)$ factors F-theory. Summarizing we can write the K\"ahler moduli as 
\begin{align}
    h^{(1,1)}(X) = 2+ \text{rk}(G) + T \, ,
\end{align}  
which is required in order to match with the gauge algebra in M-theory on $X_3$.  

From the M-theory dual, we can also deduce that the Cartan generators of the ADE singularities get enhanced to the full
adjoint valued gauge algebra that support $V=\text{dim}(ADE)$ vectors in 6D by M2 branes that wrap the respective collapsed curve. Similarly, additional hypermultiplets in the adjoint representation appear, if the D7 brane wraps a curve with genus $g>0$ \cite{Witten:1996qb}. 
Finally, we also need to know the number $h^{(2,1)}(X_3)$ of complex structure moduli. These contribute uncharged neutral hypermultiplets
\begin{align}
    H_\mathbf{1} = 1+h^{(2,1)}(X_3) \, ,
\end{align}
that are important to cancel the gravitational anomaly \eqref{eq:gravanomaly}. In addition there are massless charged hypermultiplets $H_\mathbf{R}$ that carry representations $\mathbf{R}$ under the gauge algebra $G$. From the F-theory/IIB perspective these originate from intersections of D7 brane stacks, i.e. codimension two loci, where bifundamental matter resides. However in many cases such a {\it perturbative} matter interpretation is not applicable. This is the case, whenever the brane collisions leads to {\it non-minimal} singularity in the F-theory torus, which can only be avoided upon one or multiple blow-up of the respective collision points in the base. Such configurations are called {\it superconformal matter} \cite{DelZotto:2014hpa} and their occurrence highlights the presence of tensionless strings in the theory. 

\subsubsection{SCFTs as  orbifolds} 
\label{sec:scfts-orbifolds}
 The F-theory construction gives a convenient way to geometrically engineer 6D supergravity, LSTs  \cite{Ganor:1996mu} and  SCFTs  \cite{Heckman_2015,Heckman_2019}. The latter ones are characterized via their base in terms of two-dimensional orbifolds of the type
\begin{align}
    B_{2, \text{SCFT}}= \mathbb{C}^2/\Gamma_B \, , \qquad \text{ with }\quad \Gamma_B \subset U(2) \, .
\end{align}  
This base is not required to be Calabi-Yau itself but only the elliptic total space.
When adding the elliptic fiber, one can uplift the quotient into a discrete subgroup of $SU(3)$ e.g as 
\begin{align}
  X_3=  ( \mathbb{C}^2 \times \mathbb{T}^2 )/\Gamma \, , \quad \text{with}\quad  \Gamma = ( \Gamma_B, -\det(\Gamma_B))   \subset SU(3) \, ,
\end{align}
such that $X$ is an genus-one fibered threefold.
A simple example is that of  $\Gamma_B=(e^{2 \pi i /3 },e^{2 \pi i /3 })$. Such a quotient admits a single minimal resolution in the base performed via a $(-3)$ curve  at the origin \cite{10.1007/BFb0097582,Mumford:1961vq,Morrison:2012np}. Putting a Weierstrass model over the geometry, one finds a minimal singularity of type $IV$, that is an $SU(3)$ in the F-theory context. 
This can also be seen, from the quotient perspective, by enhancing $\Gamma_B$ to 
  $\Gamma_3 = (e^{2 \pi i /3 }, e^{2 \pi i /3 },e^{2 \pi i /3 })$ which yields the singular local threefold
  \begin{align}
      X_3 = ( \mathbb{C}^2  \times \mathbb{T}^2) /\Gamma_3 \, ,
  \end{align}
  which admits three fixed codimension three fixed points in the $\mathbb{T}^2$ direction. Each fixed point can be resolved via a single resolution divisor, which yields the desired $SU(3)$ fiber over the $(-3)$ curve, which we depict as
  \begin{align}
       \overset{SU(3)}{3} \, .
  \end{align} 
  
  Several other configurations exist, with non-trivial actions on the non-compact $\mathbb{C}$ directions, in particular when $\Gamma=\mathbb{Z}_n \times \mathbb{Z}_m$ that have been explored e.g. in \cite{DelZotto:2015rca}. E.g. for $\mathbb{Z}_3 \times \mathbb{Z}_3$ with the two generators written as
  \begin{align}
   	\begin{array}{ccccc|ccccccc}
		\vartheta: && \left(\right.\;\;0\;\;\; & 1/3 & -1/3 \;\;&\; 0 & 0 & 0 & 0 & 0 & 0 \left.\right) & \\
		\omega: && \left(\right.1/3 & 0 & -1/3 \;\;&\; 0 & 0 & 0 & 0 & 0 & 0 \left.\right) & .
	\end{array}
  \end{align}
  can be thought of two local $\mathbb{Z}_3$, quotients, that lead to two $E_6$ resolutions, as described in the section before, that intersect at the origin of the F-theory base. This configuration is that of $(E_6, E_6)$ conformal matter, with tensor branch
  \begin{align}
    	[E_6]\; 1 \overset{SU(3)}{3} 1 \; [E_6]\, .
  \end{align}
  This configuration can be anticipated by noting that there are three more twisted sectors given by the combinations $\vartheta \omega$ and  $\vartheta^2 \omega$.\footnote{ $\vartheta \omega^2$ is simply the inverse of the second sector and admits no new independent fixed points.} The former one yields the $(-3)$ sector with the $SU(3)$ fiber over it while the other is empty and just yields the two $(-1)$ curves with no fiber enhancement. We will encounter such configurations in various combinations throughout the examples in the next sections. 
   
\subsection{F-theory on orbifolds}  \label{sec:FthyOnOrbifolds}
The discussion in the previous section now allows us to analyse F-theory compactifications on the geometries introduced in Section~\ref{sec:orbifoldgeometry}. 
We start with the geometry $\TT^6$ and decompose it into three two-tori. If the orbifold action respects this decomposition there are a priori three different torus-fibration structures. 
 From the M-theory perspective, each torus allows a lift to F-theory.  Note however, that not all of those lifts might be inequivalent. This   happens, if there is a permutation symmetry of the orbifold torus actions. Examples are given by six dimensional STOs with trivial affine class and orbifold generators of the same order. Turning to STOs that exhibit two generators of different order and/or non-trivial affine class destroys this symmetry and lead  to inequivalent F-theories.

In more in general however  not all STOs in the classification of \cite{Fischer_2013} can serve as an F-theory background due to the lack of at least a single torus-fibration structure. The authors of Ref.~\cite{Fischer_2013} dubbed this property factorizability. Factorizability can in general be obstructed by non-diagonal action of both the lattice as well as the point group. As in this work Abelian point groups are considered, only a lattice action which mixes the three complexified (torus-)coordinates can prevent F-theory lifts.

\subsubsection{F-theory on $\TT^4/\ZZ_3$}
\label{ssec:K3z3Ftheory}
As our first example we consider F-theory on $\TT^4/\ZZ_3$. The geometry and in particular its fixed points structure has been discussed in Section~\ref{sec-hodge-orbifold}. 
Compactifying F-theory on this four dimensional geometry yields an 8-dimensional supergravity. In fact this theory admits two equivalent F-theory lifts when taking either of the two ambient tori as the F-theory fiber. In that picture, the F-theory base is $\TT^2/\ZZ_3$, which is topologically a $\mathbb{P}^1$ with three marked points, where the F-theory torus is singular. From the resolution Figure~\ref{fig:t4-z3-fpfibration} we already know that there is an (affine) $E_6$ fiber sitting at each of the three base loci, leading to a $E_6^3$ gauge algebra. This structure in fact might be viewed as a torus compactification of IIB strings with a $\ZZ_3$ orientifold \cite{Dasgupta_1996}. An important observation made there is that the IIB axio-dilaton  is fixed to a large value, since the F-theory torus lattice is of $SU(3)$ algebra type which fixes $\tau=e^{2 \pi i/3}$. 

As noted before, this geometry is an extremal K3 which in fact is also unique. Thus, there is only one way to construct this geometry, which we do via the following Weierstrass model. In terms of the base $\mathbb{P}^1$ coordinates $u_0,u_1$ this model is given by the Weierstrass equation
\begin{align}
\label{eq:e6WSF}
 Y_2:\qquad    y^2 = x^3 +  y z^3 \left(u_0 u_1 (u_0+u_1)\right)^2 \, , 
\end{align} 
with Weierstrass functions  
\begin{align}
    (f,g,\Delta)=(0,\frac{u_0^4 u_1^4 (u_0 + u_1)^4}{4}, \frac{27  u_0^8 u_1^8 (u_0 + u_1)^8}{16}  ) \, .
\end{align}   
The Kodaira classification tells us, that there are three type $IV^*$ singularities at $u_0=0, u_1=0$ and $u_0=-u_1$, which each support an $E_6$ singularity. 
Remarkably, the above geometry admits an enhanced Mordell-Weil group of order three. This group is generated by the zero-section at  $s_0: (x,y,z)=(1,1,0)$ and the additional section  at $s_1: (x,y,z)=(0,0,1)$, which is a point of order three \cite{Aspinwall:1998xj}. As this model is unique, there is no other way of setting up an elliptic K3 with three $E_6$ fibers. This MW factor leads to a gauged center $\ZZ_3$ 1-form symmetry\cite{Aspinwall:1998xj,Mayrhofer:2014opa,Gaiotto:2014kfa}, which gives a $E_6^3/\ZZ_3$ in the 8D supergravity theory.\footnote{Those gaugings can in general lead to 8D anomalies, which are absent in cases they are realized by MW torsion of the elliptic K3 \cite{Cvetic:2020kuw} as similarly seen in the dual heterotic Narrain compactification \cite{Font:2021uyw,Fraiman:2021soq}.}

\subsubsection{F-theory on $\TT^6/\ZZ_3\times\ZZ_3-1-1$}

We move on to exemplify F-theory on a six dimensional toroidal orbifold, reviewing the results of 
\cite{hayashi2019scfts}. We do so on a $\TT^6/\ZZ_3 \times \ZZ_3$ geometry which we discussed in detail in Section~\ref{sec:geom-T6/Z3Z3-1-1}. As prepared around  equation~\eqref{eq:Z3Z311}, this geometry has three fixed point sectors of co-dimension two, that are of type $\TT^4/\ZZ_3\times \TT^2$, as well as one sector that exhibits co-dimension three fixed points.

\begin{figure}[t]
	\centering
	\includegraphics[width=0.4\linewidth]{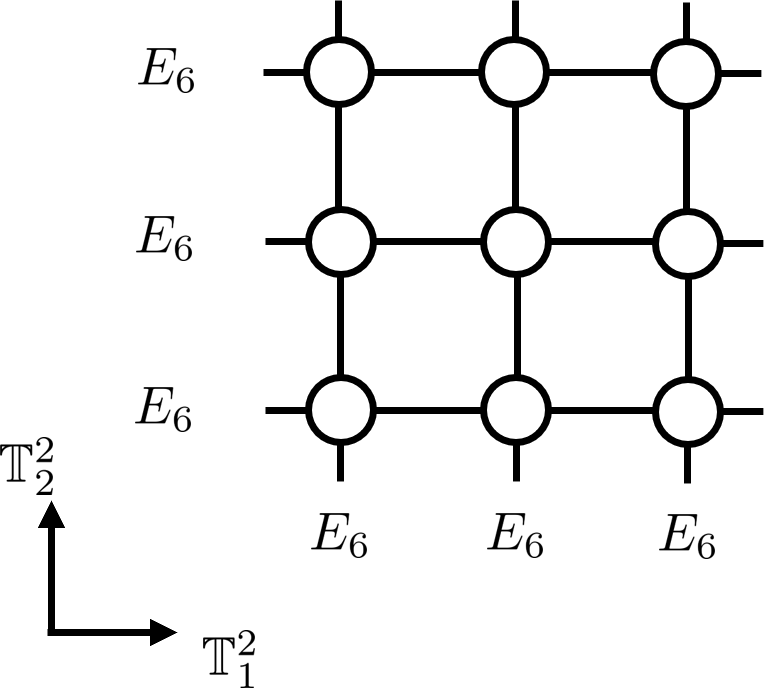}
	\caption{Depiction of the intersection structure of the six $E_6$ gauge algebra factors in the  $\TT^6/\ZZ_3\times\ZZ_3-1-1$ orbifold. At the nine intersection points, there is $(E_6, E_6)$ conformal matter, illustrated as a circle.}
	\label{fig:z3xz3intersectionstructure}
\end{figure}

We choose the third torus\footnote{Note that the geometry admits a permutation symmetry among all three tori, thus leading to equivalent F-theory lifts.} as F-theory fibre for concreteness, and torus one and two as the base part. Then the fixed points associated to the constructing elements $\vartheta$ and $\omega$ are co-dimension one singularities of the base which are the loci of D7 branes in the IIB picture. The discussion here is analogous to the $\TT^4/\ZZ_3$ orbifold, where the gauge group factors are $E_6$ for each base fixed point. Furthermore, one observes that these singularities  intersect mutually in nine points. These points are  the fixed points in the sectors associated to the $\vartheta\omega$ and $\vartheta\omega^2$ element.  As discussed in Section~\ref{sec:Ftheory}, these yields SCFT sectors namely the $(E_6, E_6)$ conformal matter whose tensor branch we recall as
\begin{align}\label{equ:E6E6SCM}
	[E_6]\; 1 \overset{SU(3)}{3} 1 \; [E_6]\, .
\end{align}
The above superconformal matter is exactly what we expect for each of the 9 $(E_6,E_6)$ collisions, that we summarized in  in Figure~\ref{fig:z3xz3intersectionstructure}.
We summarize the non-Abelian gauge algebra and the full massless spectrum of this compactification (on the tensor branch) as
\begin{align}
    \begin{array}{|c|c|} \hline G &  E_6^6 \times SU(3)^9 \\ \hline
    H  & 1 \\ \hline
    V & 6 \times \mathbf{78} + 9 \times \mathbf{8} \\ \hline
    T & 27+1 \\  \hline \end{array} \, . \nonumber
\end{align}

Note that the spectrum matches the stringy K\"ahler moduli, computed via the heterotic theory: The three untwisted K\"ahler moduli translate into volume moduli for the fiber and base plus one additional tensor. The 81 twisted K\"ahler moduli are distributed as the $36$ $E_6^6$ Cartan generators, 18 from those of $SU(3)^9$ and finally the $27$ tensors of the conformal matter points. 

This theory satisfies all 6D supergravity and gauge anomaly cancellation conditions of Section~\ref{sec:anomalies} 
which we discuss next (more details can be found in  \cite{hayashi2019scfts} and \cite{thesis}). Here we only state the results: first, the irreducible gravitational anomaly cancels provided the data above. Second, the reducible gravitational anomaly as well as mixed anomalies can be cancelled for which we first give the reduced intersection matrix
\begin{align}\label{equ:z3z3-1-1-inters}
    \Omega = \left(\begin{array}{cc}
         0 & 1 \\
         1 & 0
    \end{array}\right) \oplus \left[ \left( \begin{array}{ccc}
        -1 & 1 & 0 \\
        1 & -3 & 1 \\
        0 & 1 & -1
    \end{array} \right) \right]^{\oplus 9}\, , 
\end{align}
where the curves supporting the $E_6$ gauge group factors are in the classes of the first two components of $\Omega$, e.g. 
\begin{align}
    b_{(E_6)_1}= (1, 0, \left\{-2, -1, -1\right\}^{\oplus 3}, \{0\}^{\oplus (3\times 6)})\, .
\end{align}
The remaining nine factors correspond to the $(E_6, E_6)$ conformal matter (see equation \eqref{equ:E6E6SCM}), e.g. the $(-3)$-curve in the first factor 
\begin{align}
    b_{112}^{(-3)}=(0, 0, \{0, 1, 0\}, \{0\}^{\oplus (3\times 8)})
\end{align}
Note, that this intersection pairing is unimodular and its signature agrees with $(+, -)= (1, T)=(1, 28)$ as required.

The components of the canonical class can be found by adjunction and requiring the correct genera of the appearing curves. This results in \cite{hayashi2019scfts} in the Ansatz
\begin{align}
    a_i = \sum\nolimits_{j} (\Omega^{-1})_{ij} (2g_j - 2 - (C_j)^2)\, , 
\end{align}
where $g_j$ and $(C_j)^2$ denote the genus and self-intersection, respectively, of the $j$'th curve. This evaluates here to
\begin{align}
    a = (-2, -2, \left\{2, 1, 2\right\}^{\oplus 9})\, .
\end{align}
Using these representations as well as the intersection pairing \eqref{equ:z3z3-1-1-inters} one can show the vanishing of the anomalies reviewed in Section~\ref{sec:anomalies} and thereby verify the consistency of this 6D theory.

Finally we also remark, that the above geometry is as also consistent with a global $\mathbb{Z}_3$ torsion factor that acts diagonally on the various group factors. This is in fact also compatible with the $(E_6,E_6)/\ZZ_3$ non-simply connected conformal matter \cite{Dierigl:2020myk}. This suggests that the actual global gauge group should rather be $(E_6^6 \times SU(3)^9)/\mathbb{Z}_3$.

\begin{figure}[t]
	\centering
	\includegraphics[width=0.4\linewidth]{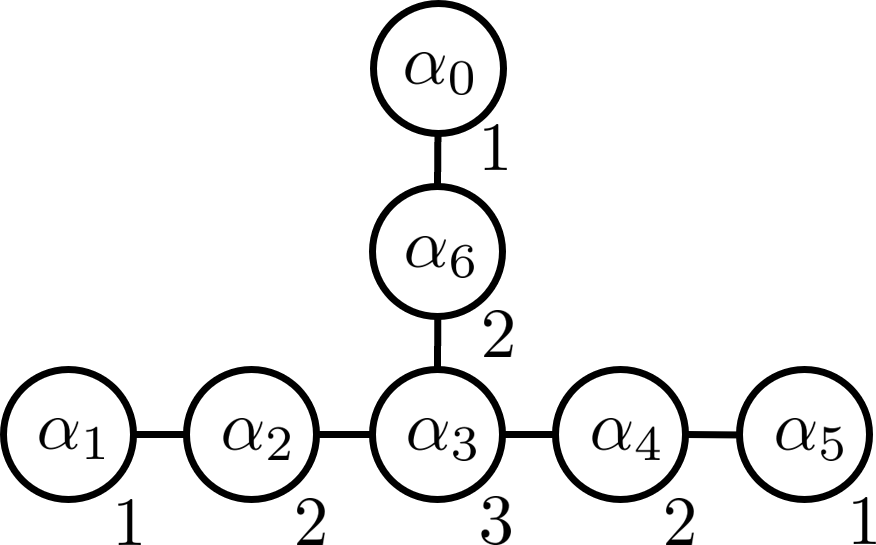}
	\caption{Depiction of the affine Dynkin diagram of $E_6$. The nodes are associates to the roots $\alpha_i$ of $E_6$ and their Kac-labels. The node associtated to $\alpha_0$ is the affine node.}
	\label{fig:e6Kac}
\end{figure}
 
Before moving on to the main part we want to comment on the unusual fiber structure that appears in the $\TT^4/\ZZ_3$, $\TT^6/\ZZ_3 \times \ZZ_3$ but also all other orbifold geometries. The main remark is, that the generic torus fiber for all orbifolds represents inner $\mathbb{P}^1$'s in the affine Dynkin diagram. In the examples here, the generic torus fiber is the unique root with Kac label 3, see Figure~\ref{fig:e6Kac}. This is at odds with degenerate elliptic fibrations, where the generic fiber is indeed given by the affine node with Kac label 1, consistent with the fact that they are intersected with the zero-section.  In this way  the presence of a zero-section marks one of these outer nodes as the affine one. However, in the orbifolds there is a $\ZZ_3$ permutation symmetry among the three fibers and hence none of them can be distinguished from the others. 

Therefore, we have only two options: First we might have no section, in which case we expect a genus-one fibration with three-sections only. The expectation for F-theory on genus-one fibrations on the other hand is, that they encode a discrete $\ZZ_n$ gauge symmetry essentially given by the order of the $n$-sections \cite{Braun:2014oya,Morrison:2014era,Anderson:2014yva,Klevers:2014bqa,Mayrhofer:2014haa,Mayrhofer:2014laa,Cvetic:2015moa,Cvetic:2018bni}
and also the review \cite{Cvetic:2018bni}.

In the case at hand, we would expect a $\ZZ_3$ gauge symmetry. This explanation however, is at odds with the fact that there is no massless matter  which could be charged under the discrete group. Moreover, there are no complex structure moduli that would allow us to un-Higgs the $\ZZ_3$ to a $U(1)$ symmetry which is in conflict with common quantum gravity lore. The second option is, that we must have three sections that are permuted by the $\ZZ_3$ symmetry. Indeed, we argued already that this model should have a $\ZZ_3$ MW torsion, which modifies the global gauge group which therefore seems to be the correct interpretation. In various other examples, that we are going to discuss in the next sections, there will be further geometric features that rather point the absence of a section and hence some $\ZZ_n$ discrete gauge symmetry in the F-theory supergravity theory.

\section{F-theory on symmetric toroidal orbifolds}\label{sec:results}

Having established the geometries of symmetric toroidal orbifolds (STOs) and the framework of F-theory in the last section, we want to present the main results of our analysis. We start by giving an overview over the STOs, and their F-theory lifts, that we have considered. Many of those geometries lead to new features that have not been present in orbifold compactifications of F-theory in the literature \cite{Morrison:1996na,Dasgupta_1996,Gopakumar1996,hayashi2019scfts}. Finally, this section serves to give a detailed discussion of three example geometries that highlight those features. 

\subsection{Summary of results}

 \subsubsection{Orbifold statistics}
 The classification of  STOs \cite{Fischer_2013} comprises orbifolds in 60 different $\QQ$–classes, that  have a holonomy group which is contained in $SU(3)$. Of these, 22 are Abelian and 38 non-Abelian. As already reviewed in Section~\ref{sec:OrbiClass}, these $\QQ$-classes can be subdivided in $\ZZ$- and affine classes, which are summarized in Table~\ref{tab:orbifoldclassification} for the Abelian subset. In total there are 162 affine classes of orbifolds with Abelian point group in the mentioned $\QQ$-classes, whereof 138 have a holonomy group which is contained in $SU(3)$ but not in $SU(2)$.

 As explained in Section~\ref{sec:FthyOnOrbifolds} not all of these geometries give sensible F-theory vacua, as the fibration structure of F-theory compactifications requires the factorizability of ``the F-theory torus". Taking into account the requirement that the group action\footnote{As noted in Section~\ref{sec:FthyOnOrbifolds}, the action of Abelian point groups always factorizes all three tori, as this defines them. In this case only a non-factorizable lattice action can obstruct F-theory lifts.} needs to factorize, of the naively $3\times 138= 414$ models only $125$, that is $\SI{30.2}{\percent}$, yield sensible F-theory compactifications. Note, that not all of these F-theory models are actually different. As encountered in Section~\ref{sec:FthyOnOrbifolds}, different choices of the F-theory torus, that is different lifts, might give the same model, if the orbifold is symmetric under exchange of these  tori.
 
 In this paper, we have considered 16 different STO geometries in various $\QQ$-classes, but trivial $\ZZ$-class and their $26$ inequivalent F-theory lifts as summarized in Table~\ref{tab:results2}. In there, we make use of the nomenclature and specification of the STO geometries and computation of stringy Hodge numbers as given in \cite{Fischer_2013}. 
 While this is a small fraction of the 125 orbifold geometries that may be used in F-theory compactifications, it allows us to initiate a study of interesting geometric features. In particular, we  explore the effect of non-trivial affine classes on the compactification of F-theory. This uncovers new features, described in more detail in the next sections, that hints at a rich structure of F-theory models beyond what is already established in the literature. 
 Let us emphasise again that we only take the first step in this exploration: to determine the consistent F-theory model for the full set of STOs,  requires a more systematic approach than we undertake here. 
\clearpage
\begin{table}[h]
	\centering
	\begin{tabular}{>{\centering\arraybackslash}m{1.2cm}||>{\centering\arraybackslash}m{1.6cm}||>{\centering\arraybackslash}m{1.6cm}|>{\centering\arraybackslash}m{4.1cm}|c|>{\centering\arraybackslash}m{1cm}|>{\centering\arraybackslash}m{3cm}}
		$\!\ZZ_2\times \ZZ_2$ & $\!\!h^{(1, 1)},h^{(2, 1)}$ & lift & $G$ & $T$ & $T_{(2,0)}$ & $H_c$ \\ 
		\hline\hline
		$-1-1$ & $51,3$ & $\!\!\TT^2_1 , \, \TT^2_2 , \, \TT^2_3$ & $\prod_{i=1}^{8}SO(8)_i$ & $17$  & - & - \\ 
		\hline
		$-1-2$ & $19,19$ & $\TT^2_1$ & $\prod_{i=1}^{8}SO(5)_i$ & $1$ & $\!8\times \mathcal{A}_1$ & $\oplus_{i=1}^8\mathbf{10}_i$  \\ 
		\cline{3-7}
		&&$\TT^2_2 , \, \TT^2_3$& $\prod_{i=1}^{2}SO(8)_i$  & $9$ & - & $ \oplus_{i=1}^2\mathbf{28}_i$\\
		\hline
		$-1-3$ & $11,11$ & $\TT^2_1$ & $\prod_{i=1}^{4}SO(5)_i$ & $1$ & $\!8\times \mathcal{A}_1$ & $\oplus_{i=1}^4\mathbf{10}_i$  \\ 
		\cline{3-7}
		&&$\TT^2_2$& $\text{\footnotesize\O}$  & $9$ & - & - \\
		\cline{3-7}
		&&$\TT^2_3$& $\prod_{i=1}^{2}SO(8)_i$  & $1$ & $\!8\times \mathcal{A}_1$ & $\oplus_{i=1}^2 \mathbf{28}_i$\\
		\hline
		$-1-4$ & $3,3$ & $\!\!\TT^2_1 , \, \TT^2_2 , \, \TT^2_3$ & $\text{\footnotesize\O}$ & $1$ & $\!8\times \mathcal{A}_1$& - \\
		\hline\hline
		
		$\!\ZZ_2\times \ZZ_4$ &&&&&& \\ 
		\hline\hline
		$-1-1$ & $61,1$ & $\TT^2_1, \, \TT^2_3$ & $\prod_{i=1}^{2}(E_7)_i\times SO(8)$\newline $\times\prod_{j=1}^{4} [(SO(7))\times Sp(1)_a\times Sp(1)_b]_j$ & $21$ & - & $\oplus_{j=1}^4[1/2(\mathbf{8}, \mathbf{2}_a)\oplus1/2(\mathbf{8}, \mathbf{2}_b)]_j$ \\ 
		\cline{3-7}
		 &  & $\TT^2_2$ & $\prod_{i=1}^{10}SO(8)_i$ & 19 & - & - \\
		\hline\hline

		$\!\ZZ_2\times \ZZ_6$ &&&&&& \\ 
		\hline\hline
		$-1-1$ & $51,3$ & $\TT^2_1, \, \TT^2_3$ & $E_8\times F_4 \times SO(8)$\newline $ \times \prod_{i=1}^{4}[G_2 \times Sp(1)]_i$ & 21 &-& $\oplus_{i=1}^4[1/2 (\mathbf{7}, \mathbf{2}) \oplus  1/2(\mathbf{1}, \mathbf{2})]_i$ \\
		\cline{3-7}
		&  & $\TT^2_2$ & $ \prod_{i=1}^{8}SO(8)_i$ & $17$ & - & - \\ 
		\hline
		$-1-2$ & $31,7$ & $\TT^2_1, \, \TT^2_3$ & $ \prod_{i=1}^{2}(F_4)_i \times \prod_{j=1}^{2}(G_2)_j $\newline $ \times \prod_{k=1}^{2}SU(3)_k$ & $13$ & - & $\mathbf{26}_1 \oplus_{j=1}^2 \mathbf{7}_j$ \\ 
		\hline\hline

		$\!\ZZ_3\times \ZZ_3$ &&&&&& \\
		\hline\hline
		$-1-1$ & $84,0$ & $\!\!\TT^2_1, \, \TT^2_2, \, \TT^2_3$ & $\prod_{i=1}^{6}(E_6)_i\times\prod_{j=1}^{6}SU(3)_j$& $ 28$ & - & - \\ 
		\hline
		$-1-2$ & $24,12$ & $\TT^2_1, \, \TT^2_2$ & $E_6\times\prod_{i=1}^{3}SU(3)_i$ & $10$ & - & $\mathbf{78}$ \\
		\cline{3-7}
		&  & $\TT^2_3$ & $\prod_{i=1}^{6}(G_2)_i$ & $10$ & - & $\oplus_{i=1}^6\mathbf{7}\oplus9\times 2\times \mathbf{1}$ \\ 
		\hline
		$-1-3$ & $18,6$ & $\TT^2_1$ & $E_6\times\prod_{i=1}^{3}SU(3)_i$ & $4$ & $\!3\times \mathcal{A}_2$ & $\mathbf{78}$ \\
		\cline{3-7}
		&  & $\TT^2_2$ & $\prod_{i=1}^{3}SU(3)_i$ & $10$ & - & - \\ 
		\cline{3-7}
		&  & $\TT^2_3$ & $\prod_{i=1}^{3}(G_2)_i$ & $10$ & - & $\oplus_{i=1}^3\mathbf{7}_i$ \\ 
		\hline
		$-1-4$ & $12,0$ & $\!\!\TT^2_1, \, \TT^2_2, \, \TT^2_3$ & $\prod_{i=1}^{3}SU(3)_i$ & $4$ &$\!3\times \mathcal{A}_2$& - \\
		\hline\hline

		$\!\ZZ_3\times \ZZ_6$ &&&&&& \\
		\hline\hline
		$-1-1$ & $73,1$ & $\TT^2_1, \, \TT^2_3$ & $E_8\times E_6 \times G_2$\newline $ \times \prod_{i=1}^{3}(F_4)_i$\newline $ \times \prod_{j=1}^{3}SU(3)_j $\newline $ \times \prod_{k=1}^{3}[G_2\times Sp(1)]_k$ & $28$ &-& $\oplus_{k=1}^3[1/2(\mathbf{7}, \mathbf{1})\oplus1/2(\mathbf{1}, \mathbf{2})]_k$ \\ 
		\hline\hline
		
	\end{tabular}
\end{table}
\clearpage
\begin{table}[h]
	\centering
	\begin{tabular}{>{\centering\arraybackslash}m{1.2cm}||>{\centering\arraybackslash}m{1.6cm}||>{\centering\arraybackslash}m{1.6cm}|>{\centering\arraybackslash}m{4.1cm}|c|>{\centering\arraybackslash}m{1cm}|>{\centering\arraybackslash}m{3cm}}

		$\!\ZZ_6\times \ZZ_6$ & $\!\!h^{(1, 1)},h^{(2, 1)}$ & lift & $G$ & $T$ & $T_{(2, 0)}$ & $H_c$ \\
		\hline\hline
		$-1-1$ & $84,0$ & $\!\!\TT^2_1 , \, \TT^2_2 , \, \TT^2_3$ & $\prod_{i=1}^{4}(E_8)_i\times \prod_{j=1}^{3}(F_4)_j$\newline $ \times\prod_{k=1}^{6}(G_2  \times Sp(1))_k $\newline $ \times SU(3) $ & $34$ &-&  $\oplus_{k=1}^6[1/2 (\mathbf{7}, \mathbf{2}) \oplus  1/2(\mathbf{1}, \mathbf{2})]_k$ \\ 
		\hline\hline

		$\!\ZZ_4\times \ZZ_4$ &&&&&& \\
		\hline\hline
		$-1-1$ & $90,0$ & $\!\!\TT^2_1 , \, \TT^2_2 , \, \TT^2_3$ & $\prod_{i=1}^4(E_7)_i\times\prod_{j=1}^{2}[SO(7)\times Sp(1)]_j$\newline $ \times\prod_{k=1}^{4}[Sp(1)_a\times SO(7)\times Sp(1)_b]_k$ & $30$ &-& $\oplus_{j=1}^2[1/2(\mathbf{8}, \mathbf{2})]_j $\newline $ \oplus_{k=1}^4 \left[1/2(\mathbf{2}_a, \mathbf{8}, \mathbf{1})\right.$\newline$\left.\oplus1/2(\mathbf{1}, \mathbf{8}, \mathbf{2}_b)\right]_k$ \\ 
		\hline\hline
		
		$\!\ZZ_6\times \ZZ_6$ &&&&&& \\
		\hline\hline
		$-1-1$ & $84,0$ & $\!\!\TT^2_1 , \, \TT^2_2, \, \TT^2_3$ & $\prod_{i=1}^4(E_8)_i\times \prod_{j=1}^3(F_4)_j$\newline $ \times\prod_{k=1}^6[G_2\times Sp(1)]_k$\newline $ \times SU(3)$ & $34$ &-& $\oplus_{k=1}^6[1/2(\mathbf{8}, \mathbf{2})]_j $ \\ 
		\hline\hline
		
		$\!\ZZ_6-II$ &&&&&& \\
		\hline\hline
		$-1-1$ & $35,11$ & $\TT^2_1$ & $ \prod_{i=1}^3(F_4)_i \times \prod_{j=1}^4(G_2)_j$ & $13$ &-& $\oplus_{i=1}^3\mathbf{26}_i\oplus_{j=1}^4\mathbf{7}_j$ \\ 
		\cline{3-7}
		&  & $\TT^2_2$ & $\prod_{i=1}^2(E_6)_i\times \prod_{j=1}^4SU(3)_j$ & $13$ &-& $\mathbf{78}_1$ \\ 
		\cline{3-7}
		&  & $\TT^2_3$ & $\prod_{i=1}^5 SO(8)_i$ & $13$ &-& $\mathbf{28}_1$ \\ 
		\hline
		
	\end{tabular}
	\caption[Results: gauge groups and gauge groups of compactifications on all considered orbifolds]{This table summarizes the massless matter content of F-theory compactifications of the various STOs.  The first two columns show the STO class and their  Hodge-numbers. The four columns to the right provide for the specified choices of F-theory fibre, the gauge groups of the theory $G$ ($\text{\footnotesize\O}$ denotes no gauge group), the number of tensor multiplets $T$ as well as the representations of the charged hyper multiplets at the tensor branch. When multiple fibers are present we highlight the $(2,0)$ tensors in the sixth column.
	The sub index of the representation specifies the factor of $G$ under which the multiplets are charged, i.e. if $G=G_1\times G_2$, then $\mathbf{R}_1$ means $(\mathbf{R}, \mathbf{1})$.  
	We also do not depict possible discrete gauge factors.
}
	\label{tab:results2}
\end{table}
\clearpage

\subsubsection{Summary of new F-theory features}\label{sec:new-features}

The STOs that we are considering throughout admit a couple of interesting geometric features which translates also into their F-theory physics. In the following we briefly summarize those which allowed us to compile Table~\ref{tab:results2}.

\begin{figure}[t!]
     \centering
     \begin{picture}(200,200)
  \put(200,0){\includegraphics[scale=0.6]{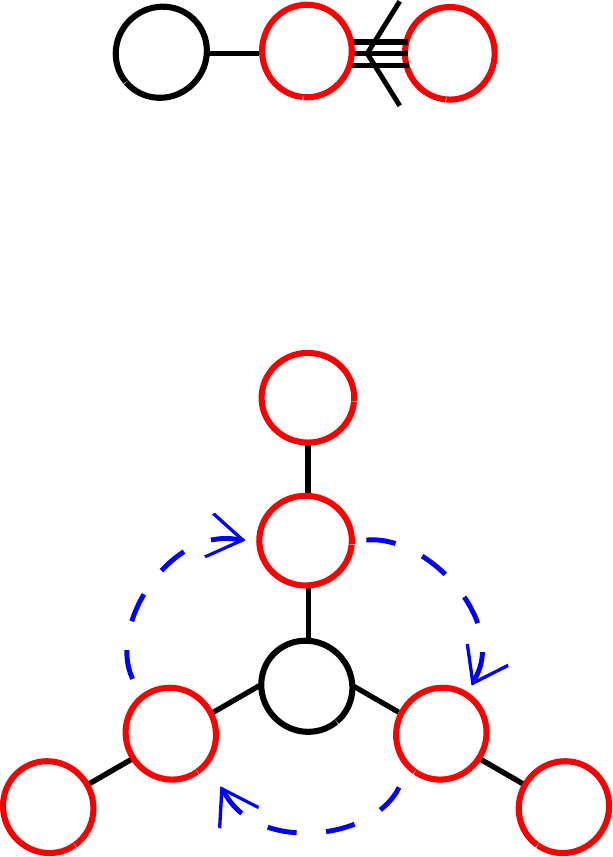}}
  \put(75,25){\includegraphics[scale=0.6]{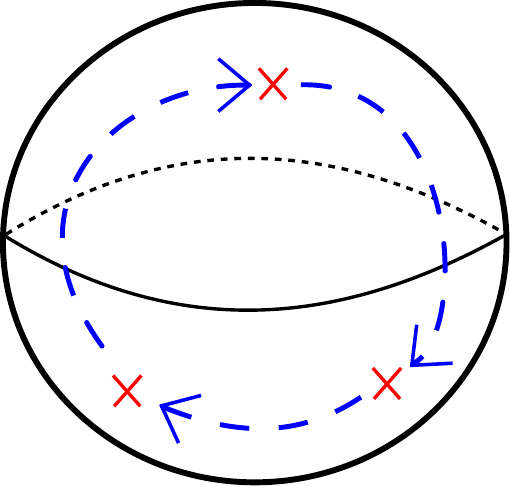}}
   \put(120,115){$z_{f_1}$}
   \put(148,30){$z_{f_2}$}
   \put(80,30){$z_{f_3}$}
   \put(247,105){\LARGE $\mathbf{\uparrow}$ }
   \put(90,80){ \textcolor{blue}{$\mathbf{\omega} $}  }
   \put(225,45){ \textcolor{blue}{$\mathbf{\omega} $ }   }
   
   \put(-120,25){\includegraphics[scale=0.25]{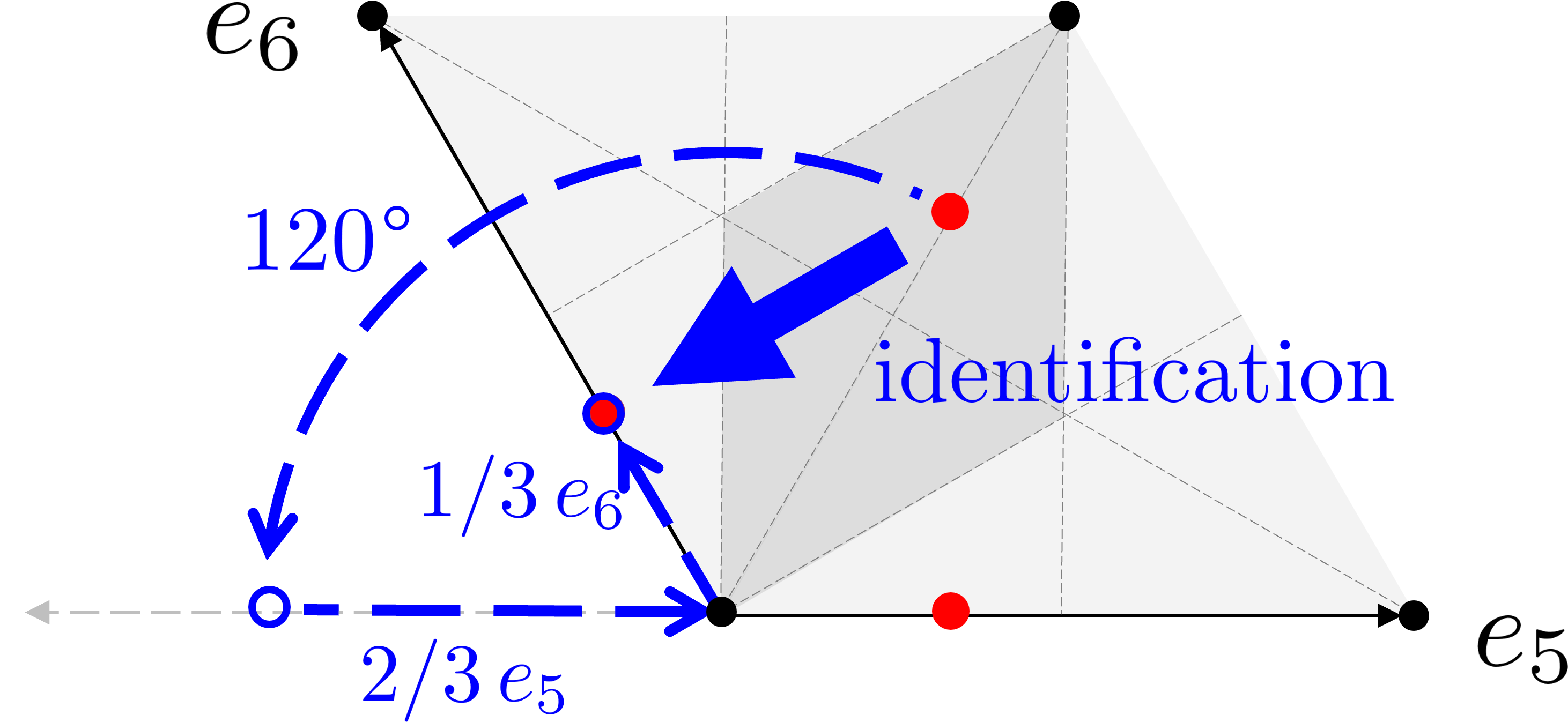}}
     \end{picture} 
     \caption{A depiction of the three $\ZZ_3$ fixed points of $\vartheta$ and their identification by the element $\omega$. In the left we show the reduction of the fundamental domain and its action on the singular $\mathbb{P}^1$. On the right we depict the action on the resolved fibers twisting the $E_6^{(1)}$ affine fiber to $D_4^{(3)}$.}
     \label{fig:E6Folding}
 \end{figure}
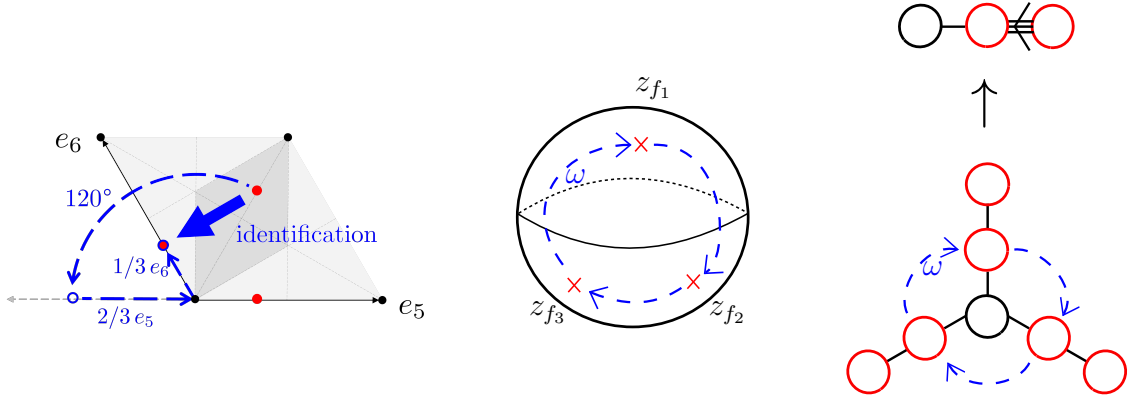

\paragraph{Multiple F-theory lifts and 5D dualities}
Most of the STOs have multiple factorizable torus structures.
In fact, since the construction descends from $\mathbb{T}^2 \times \mathbb{T}^2 \times \mathbb{T}^2$,  we have three  
F-theory lifts at our disposal whenever the orbifold action is factorizable. STOs $X$ with multiple fibration structures are interesting as they yield the same 5D theory upon circle compactification, where they are dual to M-theory on $X$.  In such theories, the distribution into 6D Vector and tensors is reshuffled and related theories  satisfy
\begin{align}
    \text{rank}(G) + T = h^{(1,1)}(X) - 2.
\end{align}
Such multiple fibration structures are a very frequent feature of
e.g. smooth CY geometries. In fact, the three fibration structures we observe, are a comparably low number, when compared e.g. with classifications in smooth CICY \cite{Anderson:2017aux} or toric hypersurface \cite{Huang:2018esr} threefolds.
This is due to the fact that the orbifold describes a CY at a particularly singular point in its K\"ahler moduli space. Indeed, we find that generically all but the three torus K\"ahler moduli are taken to zero. In the smooth geometry however, one expects to find many more linear combinations of those divisor classes
that can give a torus fiber, or in more technical terms a Kollar  divisor.
\paragraph{Identifications in fibre and base singularities} An important new class of geometries are those, that admit  additional shifts accompanied with some point group action, also known as roto-translations. The first case of such generators are those, where the shift appears in the same torus as the orbifold.  An order three example is given by  
 \begin{align}\label{equ:ex.element} 
  \begin{array}{c}
        \vartheta :
    \end{array}
    \;\;\;
	\begin{array}{ccc|ccccccc} 
		\left(\right.0\;\; & 1/3 & -1/3 \;\;&\; 0 & 0 & 0 & 0 & 2/3 & 1/3 \left.\right)   
	\end{array}\, ,
 \end{align} 
that acts on the local geometry
\begin{align}
   X_3 = (\mathbb{C} \times \mathbb{T}_2^2 \times \mathbb{T}^2_3)/\vartheta =   \mathbb{C} \times (\mathbb{T}^2_2  \times \mathbb{T}^2_3)/\vartheta    \, .
\end{align}
The above geometry admits nine fixed points in the $\mathbb{T}_2^2 \times \mathbb{T}_3^2$ directions. Focusing on $\mathbb{T}^2_3$, the three fixed points are given as
\begin{align}
    z_{f_1}= 1/3 \, e_5, \quad z_{f_2}= 1/3 \, e_6, \quad  z_{f_3}= 2/3 \, (e_5+e_6).
\end{align}
Note that the three fixed points are at different positions of the torus, due to the roto-translational element. The relevant insight is when considering another non-trivial orbifolding element given as 
  \begin{align}
  \begin{array}{c}
        \omega :
    \end{array}
    \;\;\;
	\begin{array}{ccc|ccccccc}
		\left(\right.1/3 & \;0\; & -1/3 \;\;&\; 0 & 0 & 0 & 0 & 1/3 & \,\,2/3 \left.\right) 
	\end{array}\, .
 \end{align} 
 This element leads to its own set of fixed point, but, more importantly, it also conjugates the fixed points of the the $\vartheta$ sector. In particular it acts as a cyclic translation of fixed points in the $\mathbb{T}^2_3$ direction as  
 \begin{align}
     \omega ( z_{f_i} ) = z_{f_{i+1}} \, .
 \end{align}   Without $\omega$, the resolution of those fixed points in the $\TT^2_3$ torus leads to an $E_6^{(1)}$ type fiber just as in the K3 example. The $\omega$ elements however, mods out the $\ZZ_3$ cyclic permutation symmetry of the three fixed points, which extends to a quotient by the outer automorphism of affine $E_6^{(1)}$ fiber. The folded geometry results in a $D_4^{(3)}$ twisted affine fiber as depicted in  Figure~\ref{fig:E6Folding}.
 
 Identifications of fixed points also appear e.g.
   in $\mathbb{Z}_n$ orbifolds when $n$ is non-prime and leads to fibers with non-simply laced groups. However, the free action we have encountered above always acts on (any choice of) the affine node as well and thus leads to a twisted affine type of fiber. Such actions though can not respect a zero-section and thus are associated to genus-one fibrations  \cite{Braun:2014oya}.
Following the prescription in \cite{Braun:2014oya}, such fibers are associated to a twisted version of $G_2$ in the F-theory physics. In total such fixed points come in three copies 
due to the fixed points in the $\TT_2^2$ direction. Hence in total we identify this as a $G_2^3$ configuration.  Note that this is locally, like the extremal $K3$ considered in Section~\ref{sec-hodge-orbifold} where the $E_6^3$ is reduced to $G_2^3$ by the quotient.

This geometry also admits a second fibration in the $\mathbb{T}^2_2$ direction. Here we find regular affine $E_6^{(1)}$ fiber over each of the fixed points in the $\mathbb{T}^2_3$ direction, which without the $\omega$ element   also leads to a $E_6^3$ theory. The quotient action reduces this multiplicity however by three reducing the lift to a single $E_6$ theory. Note that we have considered only a subsector of the theory, which in the full compactification is given by the $\ZZ_3 \times \ZZ_3-1-2$ orbifold, discussed in detail in Section~\ref{sec:Fthy-Z3Z3-1-2}.

\begin{figure}[t]
	\centering
	\includegraphics[width=0.7\linewidth]{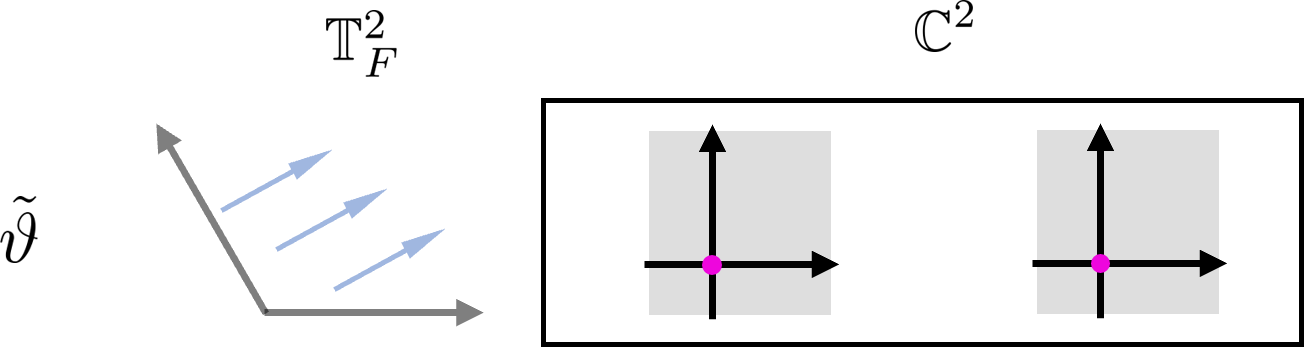} 
	\caption["Would-be" fixed point sector associated to $\tilde\vartheta$]{Illustration of a multiple fiber cause by $\tilde\vartheta$. While $\tilde\vartheta$ leads to an $\mathbb{C}^2/\ZZ_n$ orbifold base, (pink), the simultaneous translation in $\mathbb{T}^2_F$ renders the total space smooth. }
	\label{fig:z3z3-vanishing-theta-2}
\end{figure}

 \paragraph{Freely acting generators and $(2,0)$ tensors} The final class of roto-translations that is of relevance for us appears, when  point group rotations and shifts appear in different planes. An example is given by the following order three space group element   
\begin{align}
    \begin{array}{c}
        \tilde{\vartheta}:  
    \end{array}
    \;\;\;
	\begin{array}{ccc|ccccccc}
		\left(\right.0\;\; & 1/3 & -1/3\;\; &\; 2/3 & 1/3 & 0 & 0 & 0 & 0 \left.\right)  
	\end{array}\, ,
\end{align} 
which acts on the local geometry as
 \begin{align}
     X_3 = (\mathbb{T}^2  \times \mathbb{C}^2)/\tilde{\vartheta} \, .
 \end{align} 
 The important observation to make is, that 
  $\tilde{\vartheta}$ acts freely on $\mathbb{T}^2  \times \mathbb{C}^2$ and thus the total space $X_3$ is  smooth. The F-theory interpretation however is more interesting as the physical base is of the local form   $B_2 = \mathbb{C}^2/ \mathbb{Z}_3$ and hence admits a codimension two crepant orbifold singularity at the origin, see Figure~\ref{fig:z3z3-vanishing-theta-2}. This singularity however is avoided in the total space by the accompanied torus shift of $\tilde{\vartheta}$.  
  
  Fibers of this type have been discussed in the little string theory context  \cite{Bhardwaj_2016} and in compact cases in \cite{Anderson:2018heq,Anderson:2019kmx,Oehlmann:2019ohh}\footnote{See \cite{Gross97} where multiple fibers appear in the finiteness proof of elliptic threefolds. } and are called {\it multiple fibers} as the torus fiber becomes a multi-copy over the origin of the base. As the action of $\tilde{\vartheta}$ is free on the fiber, it can not respect the zero-section. As a consequence, the fibration is no longer elliptic (it does not have a section), and is only of genus-one type as discussed before. Hence again, we are led to the conclusion that we should expect a discrete gauge symmetry of the order of the quotient in F-theory \cite{Braun:2014oya}.
 However, in addition to this gauge group structure there is the important contribution of the $\mathbb{C}^2/\ZZ_n$ singularity to the spectrum. Such isolated singularities locally lead to $\mathcal{A}_{n-1}$ $(2,0)$ superconformal theories \cite{DelZotto:2014fia}. The non-trivial fiber action gauges the theory under the $\ZZ_n$ gauge symmetry.  From the perspective of anomalies, each such sector contributes like  
 \begin{align}
 \mathcal{A}_{n-1} \quad \sim \quad (1 H + 1 T)\times (n-1)  \, ,
 \end{align}
that is a neutral hyper and a tensor multiplet. It is important to remark though, that the actual tensor branch of those theories differs from a simple $A_n$ type theory by the appearance of additional discrete charged singlets over the tensors \cite{Anderson:2018heq,Anderson:2019kmx}.

For computational purposes however, those sectors contribute exactly the degrees of freedom of a regular $(2,0)$ theory as discussed above, which is sufficient to cancel all supergravity anomalies.  

 \paragraph{Discrete higher form  symmetries}
 We close this section by noting that the above geometries generically admit a rich structure of discrete symmetries as it is typical e.g. in heterotic string compactifications on such spaces. In particular, as discussed above, many orbifold geometries are genus-one fibrations, which
 hints at the possibility for non-trivial $\ZZ_n$ discrete gauged symmetries. Such symmetries include discrete 0-form symmetries but there are generically higher form discrete gauge symmetries present as well. In fact, a gauged center 1-form symmetry generated will typically be generated by a finite MW group, such as in the examples discussed in Section~\ref{subsec:z3z3-1-4} and Appendix~\ref{sec:Fthy-Z2Z2}. In the $\TT^6/\ZZ_3$ case we even find a free MW group that leads to (non-Higgsable) $U(1)$ gauge factors. Moreover, we also expect discrete  2-form gauged symmetries $G$ present. Such a discrete  symmetry $G$ appears \cite{Braun:2021sex} whenever the sublattice of collapsed divisors $\Lambda_S$ related to codimension two orbifold fix points in the base $B$ is non-primitively embedded into that of the full base $\Lambda_B$. As shown in \cite{Braun:2021sex} this can be expressed as the following torsion group, given as
 \begin{align}
     G = \text{Tor}\left( \Lambda_B / \Lambda_S    \right) \, .
 \end{align}
Generically, such a torsion element $\Gamma_B$ is present, whenever the singular base can be written as $B_\text{sing}=B/\Gamma_B$.  Physically, such 2-form symmetries couple to the light self-dual strings that are supported from D3 branes that wrap the collapsing divisors that span $\Lambda_S$. These light BPS strings exhibit a global 2-form symmetry $G_S = \text{ord}(\Gamma_B \cdot k)$ which is generically broken.
The presence of the above torsional $\Gamma_B$ element however shows that a remnant that acts diagonal on the SCFT centers is still present as a gauge symmetry.

 Since the bases we are considering here, can generically be represented as 
\begin{align}
    B = \mathbb{T}^4 / \Gamma_B \, ,
\end{align}
it follows, that we generically have a non-trivial discrete 2-form symmetry of order   $\ZZ_{\text{ord}(\Gamma_B)}$, that acts on the respective light strings. We remark, that such actions are trivial, if the respective SCFT sectors admit only trivial 2-form symmetries, as in the case of the superconformal matter, or E-string theories. This is due to the fact, that the determinant of the intersection pairing of such configurations is equal to one. For configurations such as the  $(E_6,E_6)$ conformal matter, as discussed in Section~\ref{sec:scfts-orbifolds} this can be readily satisfied. On the other hand, SCFTs that collapse to orbifold singularities do admit a 2-form symmetry, such as $\ZZ_n$ for non-Higgsable clusters on $(-n)$ curves.

\subsection{F-theory on $\ZZ_3\times\ZZ_3$ orbifolds} 
In this subsection compactifications of F-theory on several orbifolds are discussed illustrating the effects listed above. These examples generalize the ones given in Section~\ref{sec:FthyOnOrbifolds} in different ways.  First the different lifts that can be performed in the orbifold $\ZZ_3\times\ZZ_3-1-2$ are analyzed. As this is the first non-trivial affine class of  $\ZZ_3\times\ZZ_3$ orbifolds on the ordinary $SU(3)^3$ lattice, it is the natural way to generalize the $\ZZ_3\times\ZZ_3-1-1$ example of Section~\ref{sec:FthyOnOrbifolds} and of \cite{hayashi2019scfts}. One can observe some of the effects exclusive to non-trivial affine classes discussed in the previous subsection here, such as twisted affine fibres. 

As a second example we discuss the maximal (fourth) affine class of $\ZZ_3\times\ZZ_3-1$ orbifolds. This orbifold can equivalently be seen as a freely acting quotient of the single quotient orbifold $\ZZ_3-1-1$. This relation is elaborated on.  
\subsubsection{F-theory on $\ZZ_3\times\ZZ_3-1-2$}\label{sec:Fthy-Z3Z3-1-2}
\begin{figure}[t]
	\centering
	\includegraphics[width=0.7\linewidth]{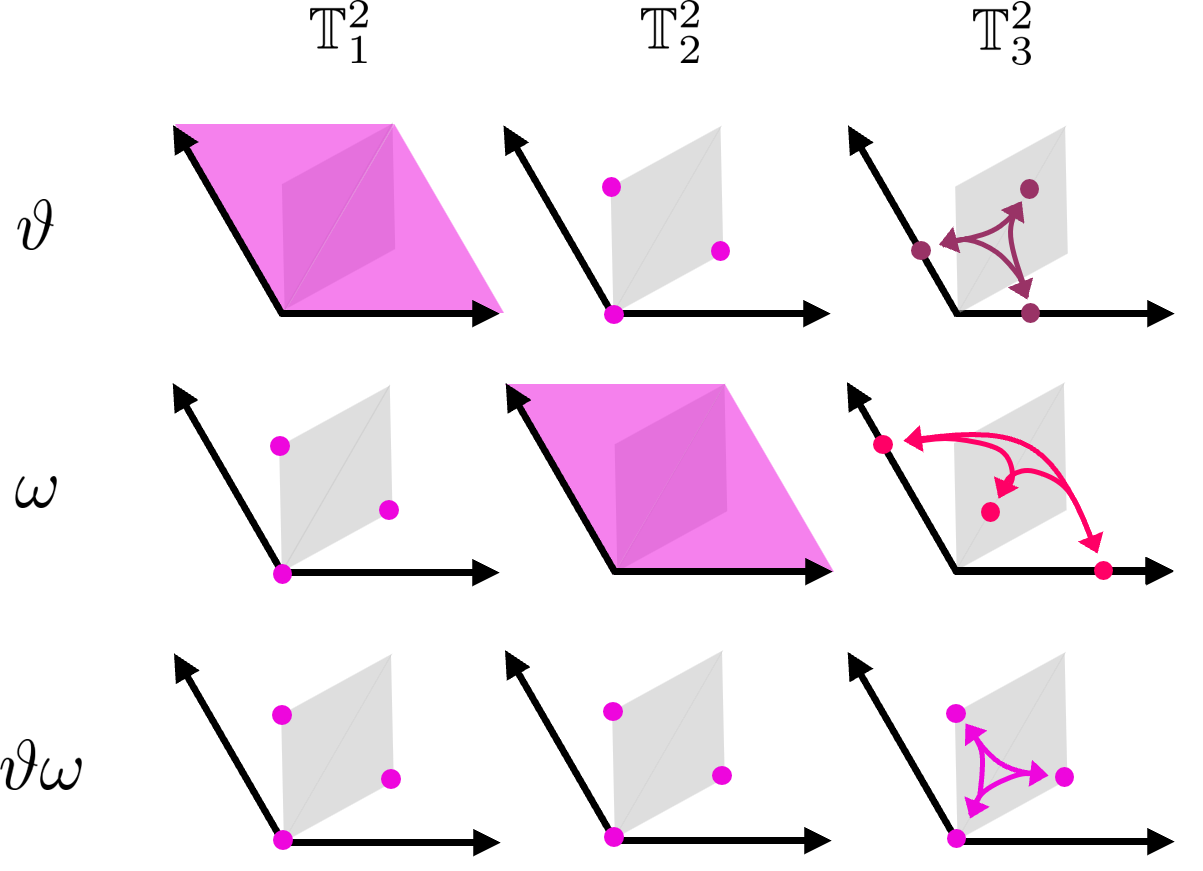}
	\caption[Fixed point structure of $\ZZ_3\times \ZZ_3-1-2$ orbifold]{Illustration of the fixed point structure of the orbifold class $\ZZ_3 \times \ZZ_3-1-2$. The fixed points are divided into three groups, according to their constructing element ($\vartheta$, $\omega$,$\vartheta\omega$), and represented by their location in the respective torus (pink, red, and purple). Arrows between fixed point locations indicate identification.}
	\label{fig:z3xz3-2-t3fpintori-1}
\end{figure}

We discuss now the first minimal generalization of the $\ZZ_3 \times \ZZ_3$ orbifold by a tranlational lattice shift into the two orbifold elements.  The two generators are now given as   \cite{Fischer_2013}
\begin{align}
    \begin{array}{c}
        \vartheta: \\\omega: 
    \end{array}
    \;\;\;
	\begin{array}{ccc|ccccccc}
		\left(\right.\;\;0\;\;\; & 1/3 & -1/3\;\; &\; 0 & 0 & 0 & 0 & 2/3 & \,\,1/3 \left.\right)  \\
		\left(\right.1/3 & 0 & -1/3 \;\;&\; 0 & 0 & 0 & 0 & 1/3 & \,\,2/3 \left.\right) 
	\end{array}\, .
\end{align}
The induced fixed points are depicted in Figure~\ref{fig:z3xz3-2-t3fpintori-1}. These yield the following stringy Hodge numbers which we can compute via the heterotic string \begin{align}
\label{eq:Hodgez3z312}
	(h^{(1, 1)}_{\text{untw.}}, h^{(1, 1)}_{\text{tw.}}) = (3, 21)\, , && 	(h^{(2, 1)}_{\text{untw.}}, h^{(2, 1)}_{\text{tw.}}) = (0, 12)\, .
\end{align} 
As the action of the orbifolding group on the three tori is not symmetric in the three tori anymore compared to the trivial affine class, there are two inequivalent F-theory lifts. These are discussed in the following.

\paragraph{F-theory lift along $\TT^2_3$}
Choosing this lift, the fractional lattice shift acts only in the F-theory torus. Thus the structure of the physical base qualitatively remains the same as in $\ZZ_3\times\ZZ_3-1-1$. There are three fixed tori in the sector associated to $\vartheta$ mutually intersecting three fixed tori in the sector associated to $\omega$. The intersections are located at the fixed points associated to $\vartheta\omega$, which are of co-dimension three in the orbifold. Examining the fibre over the fixed tori, one realizes, that the shift identifies the three fixed points, which are of order three. This is due to the fact that the fundamental domain of the geometry is reduced by a three-volution. The F-theoretic interpretation of this is a folding of the affine Dynkin diagram of $E_6$ to $G_2$ by modding its outer automorphism. 
 
Let us remark that the folding through which these $G_2$ arise, is different than that in Ref.~\cite{hayashi2019scfts}. In their work, the $G_2$ factors arise through an ordinary folding of an affine $D_4^{(1)}$ Dynkin diagram, where three nodes get identified while the affine node stays intact. Here, they emerge through the affine Dynkin diagram of $E_6$ in a folding including the affine node. Note that, as discussed in Section \ref{sec:new-features}, such a folding can only be realized, when roto-translations are present. Hence twisted affine foldings are an exclusive feature of orbifolds of non-trivial affine classe. 

The associated base curves, furthermore, have compared to the orbifold with trivial affine class, discussed in Section~\ref{sec:FthyOnOrbifolds} reduced self-intersection $(-3)$. The singularity at intersections of these curves is non-minimal and is resolved by the blow-up of one $(-1)$-curve in the base, yielding $(G_2, G_2)$ superconformal matter which is just an E-string\footnote{One might wonder, whether the freely acting element $\vartheta\omega^2$ could lead to contributions of multiple fibers as explained in Section~\ref{sec:new-features}). This is not the case, as the base fixed points are the same as in the $\vartheta\omega$ sector, which are already accounted for.}.   
To be explicit about anomaly cancellation, which is reviewed in Section~\ref{sec:anomalies}, the irreducible gravitational anomaly is cancelled 
\begin{align}
    \begin{array}{|c|c|} \hline G &  (G_2)^6 \\ \hline
    H  & 6\times \mathbf{7} + 18\times \mathbf{1} + 12+1 \\ \hline
    V & 6 \times \mathbf{14}  \\ \hline
    T & 9+1 \\  \hline \end{array} \, .\nonumber
\end{align}  

Note that the above identification of gauge group and tensors matches the $2\cdot 6 + 10 +2=24$ K\"ahler parameters in equation~\eqref{eq:Hodgez3z312}.
The origin of the different multiplets is the following. The vectors obviously are given by the adjoints of the  $G_2$ gauge algebra. The tensors are geometrically given by the nine blow-up modes at the intersections $(G_2, G_2)$ as well as the untwisted one from the $\TT^2 \times \TT^2$. Furthermore, there are different contributions to the hyper multiplets. There are twelve neutral hypers due to the complex structure moduli, the universal neutral hyper, as well as 54 charged hyper multiplets. $6\times 6=36$ of those are $\mathbf{7}$-plets of the $G_2$ gauge groups having charge dimension six and cancelling additionally the gauge anomaly on the $(-3)$-curves. The remaining $18$ hypers are discrete singlets located at the codimension two loci in the base. Note that we were required to add those additional discrete charged singlets in order cancel the gravitational anomaly. This is also motivated by the fact, that we expect a $\ZZ_3$ discrete gauge that we expect to be accompanied by the twisted affine $G_2$ fibers as those are only possible in genus-one fibrations \cite{Braun:2014oya}.

In order to show the cancellation of the reducible gravitational anomaly it is convenient to find the intersection matrix of base divisors. A priori it is given by
\begin{align}
    \tilde \Omega = (-3)^{\oplus6} \oplus (-1)^{\oplus9}\, , 
\end{align}
with appropriate off-diagonal entries denoting intersecting curves. However, due to the correspondence of the intersection matrix with the charge lattice, one requires that the signature of $\tilde\Omega$ is $(+, -)=(1, T)=(1, 10)$, and hence that its rank must be eleven. Similar as explained in \cite{hayashi2019scfts}, one in fact finds zero directions in $\tilde\Omega$ and one can find a intersection pairing with correct signature by the identification of the three fixed tori in one sector into one class as
\begin{align}
    C &= C_i + \sum\nolimits_{j} C_{ij}\, , \\
    C' &= C'_j + \sum\nolimits_{i} C_{ij}\, ,
\end{align}
where $C^{(\prime)}_{i}$, $i=1, 2, 3$, denote the curves corresponding to the fixed tori and $C_{ij}$, the resolution divisors. In the new basis the intersection pairing is block-diagonal as
\begin{align}
    \Omega = \begin{pmatrix}
    0 & 1 \\ 1 & 0
    \end{pmatrix} \oplus (-1)^{\oplus 9}\, ,
\end{align}
and all zero-eigenvalue directions are absent.
This intersection matrix fulfills all requirements such as the correct signature, unimodularity and also the gauge anomalies as well as the reducible gravitational anomaly can be verified. Therefore, one needs to find the representation of the canonical class in terms of the basis. As explained in Section~\ref{sec:FthyOnOrbifolds}, this can be found using \cite{hayashi2019scfts}
\begin{align}
    a_i = \sum\nolimits_{j} (\Omega^{-1})_{ij} (2g_j - 2 - (C_j)^2)\, , 
\end{align}
to be 
\begin{align}
    a = (-2, -2, (-1)^{\oplus 9})\, .
\end{align}
The agreement with these various constraints shows that the theory, in particular its massless spectrum, which is deduced above, actually constitutes a consistent supergravity theory in 6 dimensions. 
\paragraph{F-theory lift along $\TT^2_1 $} In order to analyze the other inequivalent lift, one chooses either the first or the second torus to be the F-theory fibre (in the following the first is chosen). As a result, the shift acts in one of the two base tori. This has in fact immense consequences on the intersection structure of the base. First of all, there is only one fixed torus left associated to $\omega$. Recall, in the orbifold $\ZZ_3\times\ZZ_3-1-1$ there were three tori fixed under $\omega$. These get identified here by the action of the additional shift. Furthermore, there were three more fixed tori associated to $\vartheta\omega^2$ that intersected the former ones. This orbifolding group element acts, however, freely here.

Beyond that there is the sector associated to $\vartheta\omega$, where three \textit{inequivalent} fixed points of co-dimension two are found in the base.  In the $\ZZ_3\times\ZZ_3-1-1$ model, these where located at the intersection points of co-dimension one loci. Due to the modified action of the generators, the co-dimension two and co-dimension loci are separated. This results in the fact, that the co-dimension one locus, which supports an $E_6$ singularity, is actually a fixed torus. As this is a curve of genus-one we thus expect the contribution of an additional adjoint charged hypermultiplet of $E_6$.
Then one notices, that in the $\vartheta$ and $\omega$ sector, we have fixed points purely located in the base that yield superconformal matter sectors. These sectors can be assembled to three $(-1)(-3)(-1)$ clusters, that host an $SU(3)$ over the middle ones.

In the following it will be verified that the above analysis in fact describes a consistent supergravity theory. From the considerations above, the  massless spectrum is given as
\begin{align*}
    \begin{array}{|c|c|} \hline G &  E_6\times SU(3)^3\, , \\ \hline
    H  & \mathbf{78}  + 12 +1 \\ \hline
    V & 6 \times \mathbf{14}+ \mathbf{78}  \\ \hline
    T & 9+1 \\  \hline \end{array} \, .
\end{align*}
Let us explain the spectrum in some more detail. As the gauge content of this theory is described by 
\begin{align}
    G=E_6\times SU(3)^3\, , 
\end{align}
there are $V=78+3\times 8=102$ vector multiplets. The number of tensor multiplets is geometrically counted by the volume moduli of the nine resolution divisors located at the three co-dimension two fixed points in the base as well as the universal tensor. Hence, 
\begin{align}
    T=9+1=10\, .
\end{align}
Moreover, there are $H=72+12+1=85$ hyper multiplets in this theory. $12+1$ of those are neutral and are due to the complex structure moduli and the universal hyper. The other $72$ are in the adjoint representation of $E_6$ (which has dimension $78$ and charge dimension $72$) and are required from gauge anomaly cancellation of a $E_6$ on a genus-one curve.

Again as above, in order to quantitatively check the cancellation of the reducible gravitational anomaly and the gauge anomalies, the intersection pairing of base divisors is studied. Here it is given by a block-diagonal combination of the transcendental factor 
\begin{align}
    \begin{pmatrix}
    0 & 1 \\ 1 & 0
    \end{pmatrix}\, , 
\end{align}
as well as three factors of the $(-1)(-3)(-1)$ chain which coincides with $(E_6,E_6)$ superconformal matter
\begin{align}
    \begin{pmatrix}
    -1 & 1 & 0 \\ 1 & -3 & 1 \\ 0 & 1 & -1
    \end{pmatrix}\, .
\end{align}
This intersection pairing, again, satisfies all conditions such as unimodularity and the correct signature. 
For  completeness, the representation of the canonical class in terms of this basis is given by
\begin{align}
    a = (-2, -2, \{2, 1, 2\}^{\oplus 3})\, . 
\end{align}
One might wonder how one recovers the curve that supports the $E_6$ gauge group. In this basis it is in fact given by 
\begin{align}
    b_{E_6} = (3, 3, (-3, -2, -3)^{\oplus 3})\, .
\end{align}
This fulfills all the properties, that one requires. Cancellation of the gauge anomaly requires that a curve that supports an $E_6$ singularity has genus-one and vanishing self intersection. By adjunction one finds that also its intersection with the canonical class must vanish. Furthermore, it cannot intersect any curve of self-intersection lower that $(-2)$, as this would yield additional superconfomal matter, which is not found here. All these properties can be verified using the intersection pairing described above. Note also, that the $E_6$ in this new basis in fact does intersect all the $(-1)$ curves of the superconformal matter points.
 
\subsubsection{F-theory on $\ZZ_3-1-1$ and $\ZZ_3\times\ZZ_3-1-4$}
In this subsection, we investigate F-theory on the $\ZZ_3$ orbifold together with $\ZZ_3 \times \ZZ_3 -1-4$ as the later one can be expressed simply as a freely acting quotient of the former one. In the F-theory context, such quotients have been discussed e.g.~in \cite{Anderson:2018heq,Anderson:2019kmx} and together with their effects on the F-theory physics.

\subsubsection*{F-theory on the $\ZZ_3$ orbifold}
We start by analyzing F-theory on the one-quotient $\ZZ_3$ orbifold.
For historical reasons, this geometry is special as it was the first that
has been discussed in the heterotic orbifold literature \cite{DIXON1985678,DIXON1986285} where the massless spectrum has been computed using the CFT methods. This geometry does in fact admit a couple of special features, which we will also encounter in the F-theory physics. 
The space group generator is given as
\begin{align}
\begin{array}{c} 
	\vartheta:
	\end{array}\;\;
	\begin{array}{ccc|cccccc} 
	\left(\right.1/3&1/3&1/3\;\; &\;0&0&0&0&0&0\left.\right)
	\end{array}\, . 
\end{align}
First we note, that the underlying $\mathbb{T}^2 \times \mathbb{T}^2 \times \mathbb{T}^2$ basis vectors $e_i$ are fixed to an $SU(3)^3$ root lattice form to admit the $\ZZ_3$ quotient symmetry, i.e. the complex structure of all ambient tori is fixed to $\tau = e^{2 \pi i/3}$. When choosing each of the three, as the F-theory torus we find that the theory is stuck at strong coupling. In fact, this theory is fully rigid, in that it does not admit any complex structure moduli. Such theories are interesting in that they do not allow to   further symmetry enhancement via unhiggsings or E-string transitions.
In this sense, such rigid theories yield endpoint theories that are {\it non-enhance-able}.

\begin{figure}[t]
	\centering
	\includegraphics[width=0.7\linewidth]{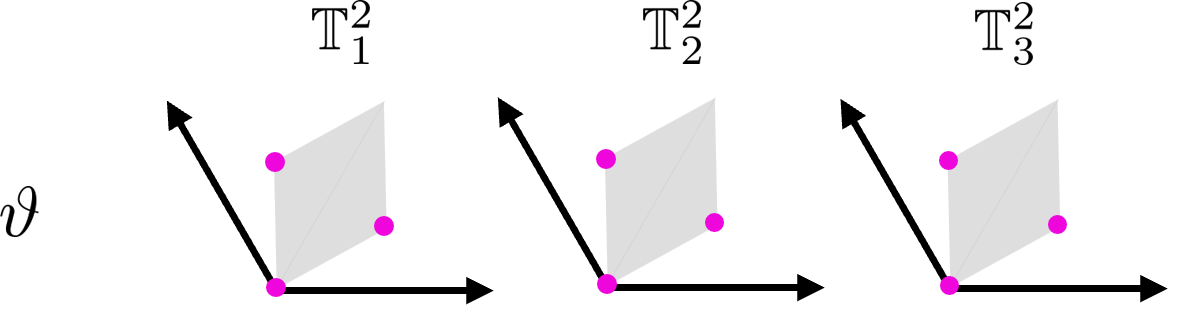}
	\caption{An illustration of the fixed points in the $\ZZ_3-1-1$ orbifold. Their location in the three tori is denoted pink.}
	\label{fig:z3fpintori}
\end{figure}

The quotient action
yields a single twisted sector with $3\times 3\times 3=27$ fixed points of co-dimension three in the total space, see Figure~\ref{fig:z3fpintori}. When computing the stringy Hodge numbers via the (2,2) heterotic string CFT, we note that something special happens for this case, as opposed to other orbifolds: The standard embedding on the heterotic side forces   the $SU(3)$ gauge bundle to be identified with the CY holonomy, which in our case is exactly the discrete $\ZZ_3$ center. 
 Hence, for this case, the heterotic gauge group is $E_6 \times SU(3)$. This is significant, as the untwisted K\"ahler moduli come in copies of $( \mathbf{27 ,3})$ bifundamentals for each ambient torus. We thus end up with nine untwisted K\"ahler modul, six more than one might expect. In addition there is one $(\mathbf{27,1})$ twisted sector field and hence K\"ahler modulus for each of the 27 fixed points. As noted before, there are no $\overline{\mathbf{27}}$, that we associated with complex structure moduli.  We summarize the stringy Hodge numbers as
\begin{align}
	(h^{(1, 1)}_{\text{untw.}}, h^{(1, 1)}_{\text{tw.}}) = (9, 27)\, , && 	(h^{(2, 1)}_{\text{untw.}}, h^{(2, 1)}_{\text{tw.}}) = (0, 0)\, .
\end{align} 

Choosing any of the three equivalent F-theory  we end up with a base of the form 
\begin{align}
B = (\mathbb{T}^2 \times \mathbb{T}^2)/\theta_B  \, , \quad \theta=(e^{i 2 \pi  i/3 },e^{i 2\pi i/3}) \, .
\end{align}
The non-crepant singularities have been analyzed already in Section~\ref{sec:Ftheory} and they are simply isolated (-3) NHCs, shrunk to the singular point. This is also consistent with the three twisted K\"ahler moduli we find for each base singularity, that yield the tensor and $SU(3)$ resolution divisors. 
We hence expect in total nine $SU(3)$ NHCs in the theory. However, now we also need to take the untwisted K\"ahler moduli into account. As noted before, these originate from the orbifold invariant K\"ahler-forms  $J \sim  J_{ab} \; d u^a \wedge d \bar{u}^{\bar{b}}$, with $a,b=1,2,3$ and $u$ being the complexified tori coordinates. The diagonal terms, with $a=b$ are the actual torus moduli. The other ones are relative rotations of these tori that appear as off-diagonal deformations in the metric.

In terms of the vectors $e_i$ of the tori \cite{Casas:1991ac} these deformations can be written as the angles
\begin{align}
\alpha_{i,j} = \frac{e_i \cdot e_j}{|e_i||e_j|} \, ,
\end{align}
with $|e_i|=R_i$.
The orbifold fixes nine of those angles and only six are free parameters. A choice for a parametrization is given as $\alpha_{1,3},\alpha_{1,4},\alpha_{1,5}, \alpha_{1,6}, \alpha_{3,6},\alpha_{3,6}$. When writing the deformations of the metric in terms of the moduli above, we get
\begin{align}
\label{eq:Z3metric}
    e_i \cdot e_j = \left(\begin{array}{cccccc} 
    R_1^2 & -\frac12 R_1^2 & R_1 R_3 \alpha_{1,3} &  R_1 R_3 \alpha_{1,4}&   R_1 R_5 \alpha_{1,5} &   R_1 R_5 \alpha_{1,6} \\  
    -\frac12 R_1^2& R_1^2  & R_1 R_3 \alpha_{2,3} &  R_1 R_3 \alpha_{1,3}&   R_1 R_5 \alpha_{2,5} &   R_1 R_5 \alpha_{1,5}    \\ \cline{3-6}
    R_1 R_3 \alpha_{1,3} &  R_1 R_3 \alpha_{2,3}  & \multicolumn{ 1}{|c}{R_3^2} & -\frac12 R_3^2&  R_3 R_5 \alpha_{3,5}&   R_3 R_5 \alpha_{3,6} \\   
        R_1 R_3 \alpha_{1,4} &R_1 R_3 \alpha_{1,3} &   \multicolumn{ 1}{|c}{-\frac12 R_3^2} & R_3^2  &  R_3 R_5 \alpha_{4,5}&   R_3 R_5 \alpha_{3,5} \\ 
         R_1 R_5 \alpha_{1,5} &  R_1 R_5 \alpha_{2 ,5 }&  \multicolumn{ 1}{|c}{ R_3 R_5 \alpha_{3,5} }&   R_3 R_5 \alpha_{4,5} &  R_5^2 & -\frac12 R_5^2 \\  
       R_1 R_5 \alpha_{1,6} &  R_1 R_5 \alpha_{1 ,6 }& \multicolumn{ 1}{|c}{  R_3 R_5 \alpha_{3,6}} &   R_3 R_5 \alpha_{3,5} & -\frac12 R_5^2& R_5^2  
    \end{array}  \right)\, .
\end{align}
In the depiction above, we have singled out the metric deformations along the first torus direction, as this is going to be our F-theory torus choice\footnote{The $\alpha_{2,3},\alpha_{2,4},\alpha_{2,5}$ can be rewritten in terms of other basis angles using $\alpha_{i,j}=-(\alpha_{i+1,j} +\alpha_{i,j+1})$. }. Here we find, when taking the F-theory lift, that is $R_1 \rightarrow \infty$, we decouple the four off-diagonal moduli   $\alpha_{1,3},\alpha_{1,4},\alpha_{1,5},\alpha_{1,6}$. Hence these moduli must be associated to horizontal divisors, and as they do not correspond to codimension one (singular) fibers, they must correspond to the K\"ahler classes that generate a rank four free Mordell-Weil group and thus a $U(1)^4$ gauge group in F-theory.

On the other hand, there are still the four relative moduli $\alpha_{3,5},\alpha_{3,6}$ of the two base $(\mathbb{T}^2 \times \TT^2)/\theta_B$ that are  dual to other base divisors and hence yield two additional tensor multiplets.
 When putting everything together, the F-theory gauge group is given as
\begin{align*}
    \begin{array}{|c|c|} \hline G &  SU(3)^9 \times  U(1)^4 \\ \hline
    H  & 1 \\ \hline
    V & 9 \times \mathbf{8} + 4 \times \mathbf{1} \\ \hline
    T & 9+3 \\  \hline \end{array} \, ,
\end{align*}
consistent with the gravitational anomaly. 

However, it remains an open task to find the complete consistent intersection matrix of the resolved base. In this structure, there will be nine curves of self intersection $(-3)$, that do not intersect each other, as well as the transcendental factor. The remaining two directions associated to the two off-diagonal K\"ahler-moduli of the fibre have negative signature (as the signature of the entire intersection matrix is supposed to be $(+, -)=(1, T)=(1, 12)$) and are supposed to intersect the other divisors in a manner, such that the resulting intersection matrix is unimodular\footnote{A resolution of this orbifold in terms of a GLSM construction can be found in \cite{Blaszczyk:2011hs}, which however is limited to the diagonal K\"ahler parameters only. In that resolution, the fiber is described as a generic cubic, that is a genus-one fibration with three-sections only.}. While we were not able to show this unimodularity, we are still confident that the analysis of the massless spectrum of F-theory on this orbifold is valid due to the reasons mentioned above.

We close by commenting on the various additional symmetries that are present. First we have noted already that the symmetry of the fixed points suggests a gauged 1-form symmetry, as in the K3 example, which implies a global $SU(3)^9/\ZZ_3$ global gauge group.  Moreover,
following the general arguments of Section~\ref{sec:new-features}, the F-theory lift admits a $\ZZ_3$ discrete 2-form gauge symmetry. As observed the nine collapsed fixed points of the $\TT^2\times \TT^2/\ZZ_3$ base
all admit self-intersection $-3$ and have hence a $\ZZ_3$ 2-form global symmetry. The diagonal $\ZZ_3$ action of the global $\ZZ_3^9$ global symmetry of the SCFT sectors.

\subsubsection*{F-theory on the $\ZZ_3\times\ZZ_3-1-4$ orbifold}
\label{subsec:z3z3-1-4}
Having established the F-theory physics of the $\ZZ_3$ orbifold before we are adding  another $\ZZ_3$ quotient factor that acts freely on the geometry and  coincides with the $\ZZ_3\times\ZZ_3-1-4 $ geometry. In order to see this, the latter geometry is discussed briefly. This STO is in the same $\QQ$- and $\ZZ$-class as the ordinary $\ZZ_3\times\ZZ_3 $ orbifold considered for example in \cite{hayashi2019scfts} and the $\ZZ_3\times\ZZ_3-1-2$ in Section~\ref{sec:results}. This geometry however admits maximal affine-class, that is exhibits the maximal amount of fractional lattice shifts. 
The two generators   can be chosen to be\footnote{This reproduces the group generators in the form as used in \cite{Fischer_2013}, as $\vartheta=\tilde{\vartheta} \tilde{\omega}$ and $\omega=\tilde{\vartheta}$, with
\begin{align}
\begin{array}{c}  
	\,
	\tilde{\vartheta}:\\
	\tilde{\omega} :
	\end{array}
	\;\;
	\begin{array}{ccc|cccccc} 
	\left(\right.\;\,0\,\;\;\;\;&1/3&-1/3\;\; &\;2/3&1/3&2/3&1/3&1/3&2/3\left.\right)\\
	\left(\right.1/3\;\;\;&0&-1/3\;\; &\;2/3&1/3&2/3&1/3&1/3&2/3\left.\right)
	\end{array}\, .
\end{align}} 
\begin{align}
\begin{array}{c} 
	\vartheta:\\
	\omega:
	\end{array}
	\;\;
	\begin{array}{ccc|cccccc} 
	\left(\right.1/3\;\;\;&1/3&1/3\;\; &\;0&0&0&0&0&\;\,0\;\,\left.\right)\\
	\left(\right.\;\,0\,\;\;\;\;&1/3&-1/3\;\; &\;2/3&1/3&2/3&1/3&1/3&2/3\left.\right)
	\end{array}\, .
\end{align}
The first factor is as in the $\ZZ_3$ orbifold and the second   $\omega$ realizes the free action.
Using the free action, we can relate indices in the geometry: If the action is free, indices like Euler numbers get simply divided by the order of the quotient. Hence in our case we have  
\begin{align}
     \chi_{\ZZ_3\times\ZZ_3 }  = \frac{\chi_{\ZZ_3 }}{3}\, .
\end{align}
and therefore relate also the stringy Hodge numbers to
\begin{align}
	(h^{(1, 1)}_{\text{untw.}}, h^{(1, 1)}_{\text{tw.}}) = (3, 9)\, , && 	(h^{(2, 1)}_{\text{untw.}}, h^{(2, 1)}_{\text{tw.}}) = (0, 0)\, .
\end{align} 

The above Hodge numbers also agree with the spectrum in the heterotic (2,2) CFT.
\begin{figure}[t]
	\centering
	\includegraphics[width=0.7\linewidth]{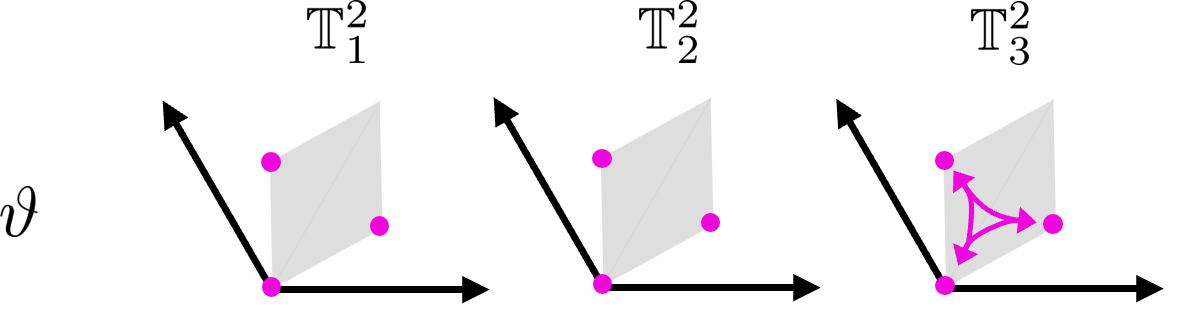}
	\caption{An illustration of the fixed points in the $\ZZ_3\times\ZZ_3-1-4$ orbifold. Their location in the three tori is denoted pink and identification is indicated by arrows.}
	\label{fig:z3xz3-4fpintori}
\end{figure}
The free generator $\omega$ relates all of the 27 fixed points and groups them into nine inequivalent ones as depicted in Figure~\ref{fig:z3xz3-4fpintori} and also removes six of the off-diagonal untwisted K\"ahler moduli leaving only the three ambient tori radii.  Lets now discuss the F-theory physics of the resulting geometry. First, there are only three $(-3)$ $SU(3)$ NHCs cluster in the base left. With respect to the untwisted moduli, we find that the additional quotient $\omega$ fixes all residual off-diagonal K\"ahler modul $\alpha_{i,j}$ in \eqref{eq:Z3metric}. Thus the quotient removes all four $U(1)$ factors and two tensors of the base. Before we can summarize the spectrum we note, that the $\ZZ_3 \times \mathbb{Z}_3-1-1-4$ is non-simply connected as a manifold\footnote{This property is also preserved in the resolution phase.}. Hence from the considerations of Section~\ref{sec:new-features} and \cite{Anderson:2018heq,Anderson:2019kmx} we expect  $\omega$ to lead to multiple fibers. Indeed, when viewed from an F-theory perspective, the $\omega$ element acts as an additional $\mathbb{Z}_3$ orbifold in the (singular) base, and while it identifies a couple of fixed points in the $\vartheta$ sector it also introduces three new $\ZZ_3$ fixed points in the base. Those fixed points count as three $\mathcal{A}_2$ theories with local $\mathcal{N}= (2,0)$ supersymmetry and thus, like $3 \times 2\cdot (H+T)$-plets. Note again, that those fixed points appear only in the base geometry but not in the total space and hence admit no associated $h^{(1,1)}$  contributions.

We are now in the position to summarize the massless F-theory spectrum to 
\begin{align*} 
       \begin{array}{|c|c|} \hline G & SU(3)^3 \\ \hline
    H  & 1 \\ \hline
    V & 3 \times \mathbf{8}   \\ \hline
    T & 3+1  \\  \hline
     T_{2,0} & 3 \times \mathcal{A}_2  \\  \hline\end{array} \, .
\end{align*}
The above spectrum spectrum is consistent with the gravitational anomaly, which notably requires the additional fixed point contribution in the base. Similar, as in the covering geometry, we expect the various curves to be glued together to lead to an uni-modular intersection pairing.

Now we can also discuss the various discrete symmetries in the theory. As discussed before, the existence of a multiple fiber is a smoking gun for a genus-one fibration. Indeed, compatibility requires to have (at least) a three-section and thus it follows that there is a discrete $\ZZ_3$ 0-form gauge symmetry. Again, there are no massless charged hypermulitplets in the theory, but the $(2,0)$ SCFT sector as explicitly discussed in \cite{Anderson:2018heq}. In addition we find, that the base is a quotient of the form
\begin{align}
B_2 =    (\mathbb{T}^2 \times \mathbb{T}^2)/(\vartheta_B \times \omega_B) \, .
\end{align}
Indeed, this implies a $\mathbb{Z}_3 \times \ZZ_3$ discrete 2-form gauge symmetry that acts diagonally on the $\ZZ_3^6$ global 2-form symmetries of the $(-3)^3 \times \mathcal{A}_2^3$ SCFT sectors.  
\section{Summary and conclusion} \label{sec:conclusion} 
Symmetric toroidal orbifold geometries are distinguished loci in the landscape of Calabi-Yau geometries, in that they are essentially flat, apart from the orbifold fixed points. This simplicity makes the geometries also very attractive for F-theory compactifications. Moreover, all six-dimensional symmetric toroidal orbifolds have been classified rather recently \cite{Fischer_2013} and therefore allows us to initiate a systematic study of F-theory on these geometries.  

Our analysis highlights a couple of significant features: First STOs admit up to three inequivalent torus fibrations. This allows to obtain multiple different F-theory lifts from the same geometry that yield the same M-theory dual. Second, in case with non-trivial affine classes the orbifold rotation is accompanied by fractional lattice shifts,  so called   roto-translations. Those shifts identify fixed points, which lead to twisted affine fibers in the F-theory torus and/or reduced amounts of SCFTs when they act in the base. Moreover, they can lead to exotic SCFT sectors supported by multiple-fibers.
In particular we find that the bases, that generically are of $\mathbb{T}^4/\Gamma$ type, can host isolated superconformal theories at the fixed points. These superconformal points include in particular non-Higgsable clusters at the origin of their tensor branch. These cluster admit, by definition, no flavor symmetry. Our study enlarges the class of configurations that have been explored in \cite{hayashi2019scfts}, where the SCFT sectors were always of superconformal matter type, that  exhibit non-trivial (gauged) flavor symmetry. 

Third,  we find that the 
IIB axio-dilaton is constant and fixed to strong coupling. In fact, many of the geometries are completely rigid,
as they do not admit any complex structure deformations. 
 The resulting supergravity can therefore be viewed as an {\it endpoint theory} that cannot admit further gauge or SCFT enhancements.
Fourth, orbifolds admit a vast amount of discrete symmetries
that impact the underlying effective string compactification, such as enhanced flavor or R-symmetries in the heterotic case. Correspondingly we also find enhanced discrete gauge symmetries in F-theory on the same spaces. These in particular include 0-, 1-, and 2-form discrete gauge symmetries.

The heterotic theories were very useful in various other ways for this works as well. The $(2,2)$ worldsheet CFT 
in \cite{Fischer_2013} helped to obtain the stringy Hodge numbers of the orbifold geometries. We  required those in order to show 6D supergravity anomaly cancellation as well as to cross-check the gauge group rank and number of tensor multiplets.  

While this work initiates a complete classification of F-theory on symmetric toroidal orbifolds, various questions and directions for future work remain. This includes in particular a better understanding of theories with twisted affine fibers and a complete derivation of the anomaly lattices for some of the geometries considered here.  
The next goal is then the completion of the study of F-theory on Abelian orbifolds for all affine-classes. 

To complete our study we would then turn to the new class of non-Abelian orbifolds and their F-theory lifts which 
we expect to lead to various other potentially more exotic theories. All stringy Hodge numbers for all 331 non-Abelian
orbifolds have been computed \cite{Fischer:2013qza} as well, such that we can simply run our program there too. Moreover we have left 
an analysis of the little string degenerations, that are readily obtained from decompactifying one of the base tori, for future work. This direction is particularly appealing, as the various torus-fibration structures which relate T-dual little string theories, are directly build into the STOs.  

This program on the other hand, can further be generalized to  toroidal four- or even five-folds which, however, are not classified so far. F-theory on those spaces lead to four and two dimensional theories. Those geometries admit a couple of additional interesting features, such as terminal
singularities that do not exhibit resolutions \cite{Arras:2016evy} or deformations.
The heterotic world sheet CFT however is still well defined, such that we can compute the corresponding massless states at the fixed points, that correspond to their deformation spaces in the standard embedding \cite{Font:2004et}. In 4D we would then expect to find supergravity theories coupled to the (local) $\mathcal{N}=3$ theories constructed in \cite{Garcia-Etxebarria:2015wns,Garcia-Etxebarria:2016erx}.

\section*{Acknowledgements}
We thank  Fabian Ruehle and Damian Kaloni Mayorga Pena for discussions and collaboration on related work. We further thank Michele del Zotto and Patrick Vaudrevange for comments and valuable discussions.  
F.B.K wants to thank Raimar Wulkenhaar for supporting this project.
The work of F.B.K. is financed by the Research Training Group 2149 funded by the Deutsche Forschungsgemeinschaft as well as supported by the Cluster of Excellence Mathematics Münster\footnote{Funded by the Deutsche Forschungsgemeinschaft (DFG, German Research Foundation) –
Project-ID 427320536 – SFB 1442.}. 
The research of M.L. is financed by Vetenskapsradet under grant number 2020-03230.
The work of P.K.O. is supported by a grant of the Carl Trygger Foundation for Scientific Research and the European Research Council (ERC) under the European Union’s Horizon 2020 research
and innovation programme (grant agreement No. 851931).   

\appendix

\section{F-theory on $\ZZ_2\times\ZZ_2$ orbifolds}\label{sec:Fthy-Z2Z2}
In order to illustrate our analysis as well as some issues that still remain, we provide the analysis of $\ZZ_2\times\ZZ_2$ orbifolds on a lattice $\Lambda=SU(2)^6$. New features that are described in the main section are also found here as well as challenges for future studies.

\subsection*{$\ZZ_2\times\ZZ_2-1-1$ orbifold}
\begin{figure}[t]
	\centering
	\includegraphics[width=0.6\linewidth]{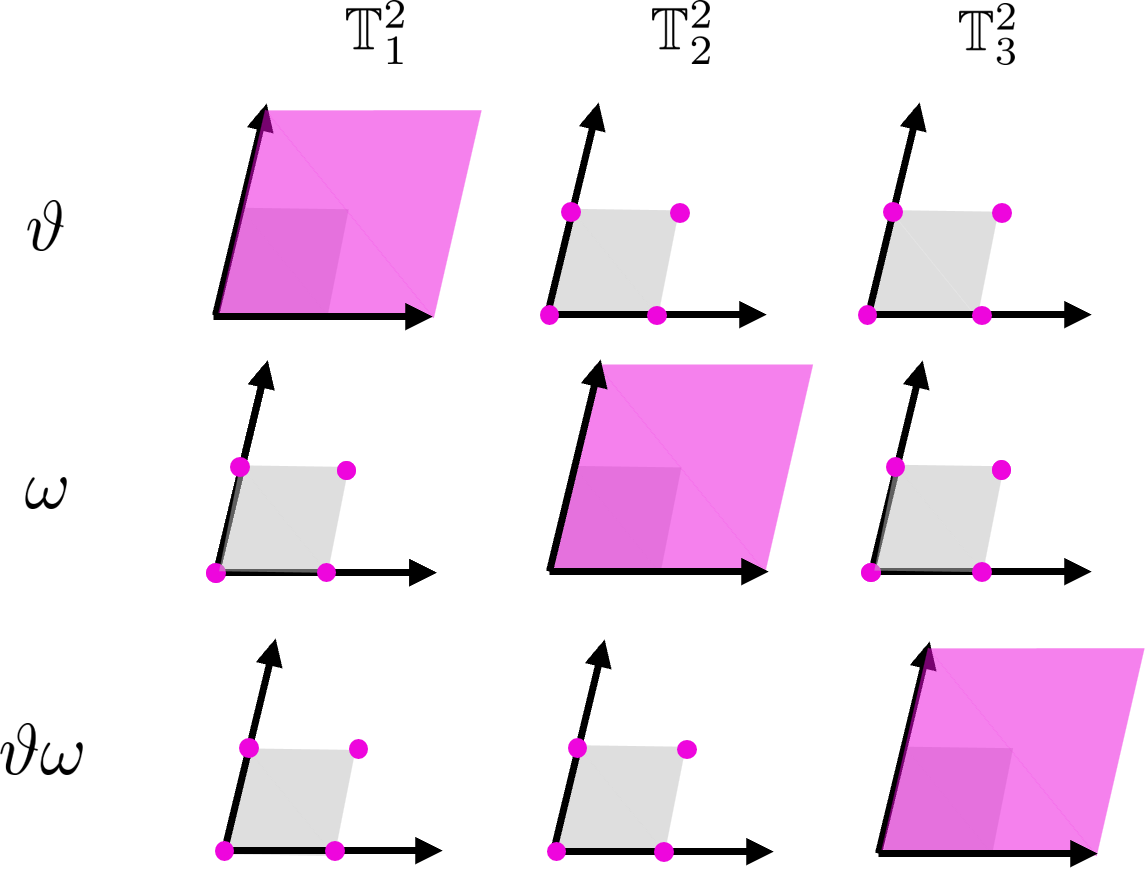}
	\caption{An illustration of the fixed points (pink) in the $\ZZ_2\times\ZZ_2-1-1$ orbifold.}
	\label{fig:z2z2-1-1-fpintori}
\end{figure}

In order to be complete and for later reference, F-theory on the standard $\ZZ_2\times\ZZ_2$ orbifold, that is the trivial affine class is very briefly reviewed here. The complete analysis can be found in \cite{hayashi2019scfts}. \\
The action of the two generators of the space group
\begin{align}
    \begin{array}{c}
        \vartheta:\\\omega:
    \end{array}
    \;\;\;
	\begin{array}{ccc|ccccccc}
		\left(\right.\;\;0\;\;\; & 1/2 & -1/2 \;\;&\; 0 & 0 & 0 & 0 & 0 & 0 \left.\right)  \\
		\left(\right.1/2 & 0 & -1/2 \;\;&\; 0 & 0 & 0 & 0 & 0 & 0 \left.\right) 
	\end{array}\,  
\end{align}
gives rise to three fixed point sectors, each containing $16$ fixed tori (see Figure~\ref{fig:z2z2-1-1-fpintori}). The stringy Hodge numbers are given as
\begin{align}
	(h^{(1, 1)}_{\text{untw.}}, h^{(1, 1)}_{\text{tw.}}) = (3, 48)\, , && 	(h^{(2, 1)}_{\text{untw.}}, h^{(2, 1)}_{\text{tw.}}) = (3, 0)\, .
\end{align}
As this structure is symmetric in the three tori, there is just one inequivalent lift. In this lift one finds $4+4$ fixed tori in the base distributed among two sectors mutually intersecting in the the fixed points of co-dimension two in the third sector. The F-theoretic interpretation of degenerated fibre over the fixed tori yields $SO(8)$ gauge group factors. The intersection of these base curves hence gives rise to $(SO(8), SO(8))$ superconformal matter. These singular loci of co-dimension one in the base are resolved by the blow-up of one $(-1)$-curve in the base. 

The massless matter spectrum of theory is summarized as follows
\begin{align}
    \begin{array}{|c|c|} 
    \hline 
    G & (SO(8))^4\times(SO(8))^4 \\ \hline
    H  & 3+1 \\ \hline
    V & (4+4)\times \mathbf{28}   \\ \hline
    T & 17  \\  \hline
    \end{array} \, .\nonumber
\end{align}
As there are no charged hyper multiplets, the total number of hyper multiplets is given by the $H=h^{(2, 1)}+1=3+1=4$. The $17$ tensors can be explained by the volume moduli of the exceptional divisors at the $(SO(8), SO(8))$ superconformal matter loci and the universal mode. 

The resulting intersection pairing, described in detail in\cite{hayashi2019scfts}, yields cancellation of all anomalies and hence a consistent theory.
Before moving on to the next theory, we also want to remark here, that we can view the geometry as the collision of two K3s that come from the Sen configuration of $\TT^2 \times \TT^2 /\ZZ_2$ that yield an $SO(8)^4$ configuration each. These K3s are the unique configurations 2147 in \cite{2005math......5140S} that admit a $\ZZ_2 \times \ZZ_2$ MW group. Hence we expect that our configuration admits a $SO(8)^8/\ZZ_2 \times \ZZ_2$ form. This is again compatible with the supreconformal matter configurations found as well \cite{Dierigl:2020myk}. Those considerations are again consistent with the presence of the four order two points that sit at at the precise locations, where the orbifold fixed points are.

\subsection*{$\ZZ_2\times\ZZ_2-1-2$ orbifold}
\begin{figure}[t]
	\centering
	\includegraphics[width=0.6\linewidth]{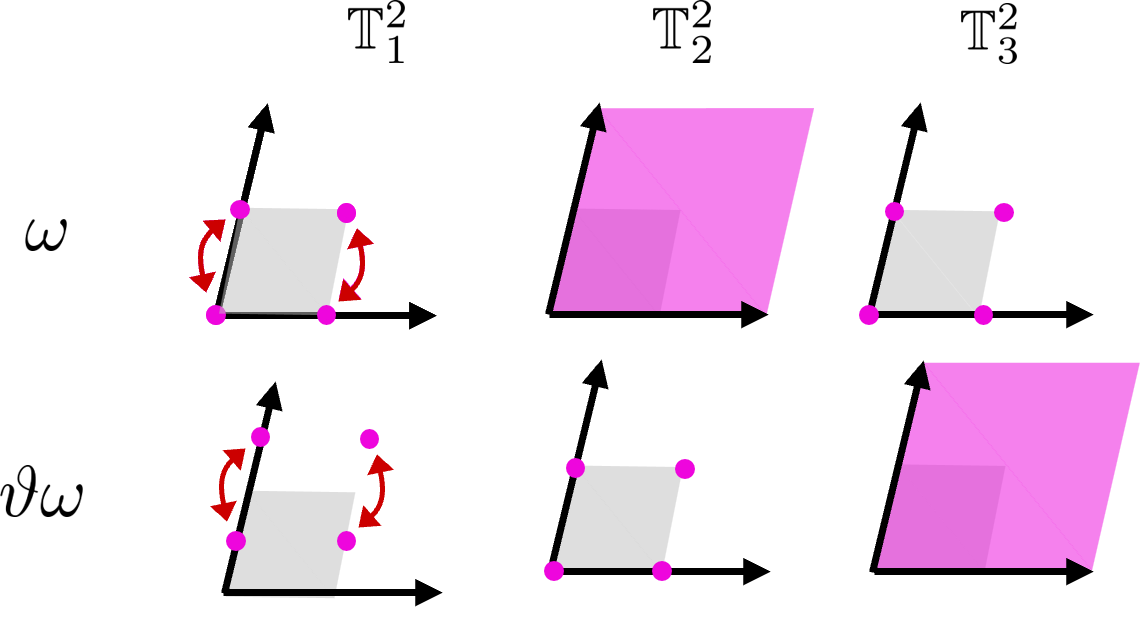}
	\caption{An illustration of the fixed points (pink) in the $\ZZ_2\times\ZZ_2-1-2$ orbifold.}
	\label{fig:Z2xZ2-1-2_FPinTori_2}
\end{figure}

Turning to the fist non-trivial affine class, a shift is introduced into the generator $\vartheta$ resulting in 
\begin{align}
    \begin{array}{c}
        \vartheta:\\\omega:
    \end{array}
    \;\;\;
	\begin{array}{ccc|ccccccc}
		\left(\right.\;\;0\;\;\; & 1/2 & -1/2 \;\;&\; 0 & 1/2 & 0 & 0 & 0 & 0 \left.\right)  \\
		\left(\right.1/2 & 0 & -1/2 \;\;&\; 0 & 0 & 0 & 0 & 0 & 0 \left.\right) 
	\end{array}\, .
\end{align} 
The fixed point structure is depicted in Figure~\ref{fig:Z2xZ2-1-2_FPinTori_2}. 
\begin{figure}[t]
	\centering
	\includegraphics[width=0.6\linewidth]{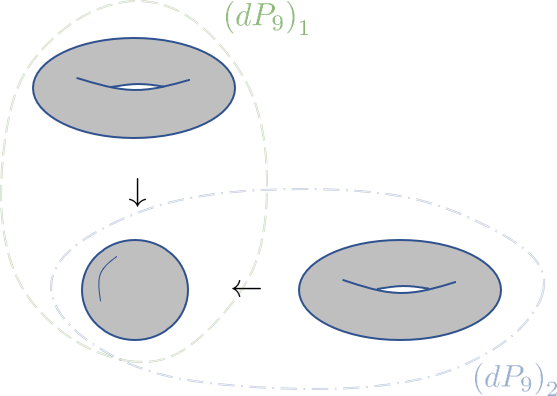}
	\caption{Illustration of the Schoen manifold as the fibre product of two del Pezzo surfaces $dP_9$.}
	\label{fig:Schoen_dP9}
\end{figure} 
As already indicated by the Hodge numbers 
\begin{align}
	(h^{(1, 1)}_{\text{untw.}}, h^{(1, 1)}_{\text{tw.}}) = (3, 16)\, , && 	(h^{(2, 1)}_{\text{untw.}}, h^{(2, 1)}_{\text{tw.}}) = (3, 16)\, ,
\end{align}
this orbifold is in fact a singular limit of the Schoen manifold \cite{Schoen1988}. This geometry has been analysed in  the heterotic orbifold context in \cite{nibbelink2013schoen} as well as in the F-theory context in \cite{Gopakumar1996}. 

The Schoen manifold is a fibre product of two rational elliptic surfaces $S_1$ and $S_2$. The $S_i$ are described by a torus fibration $\pi_i: \; S_i \to \PP^1$ such that 
\begin{align}\label{equ:Schoen-prod}
    X_{Schoen} = \{(x_1, x_2)\in S_1\times S_2: \; \pi_1(x_1)=\pi_2(x_2) \}\, .
\end{align}
This structure is illustrated in Figure~\ref{fig:Schoen_dP9}. As a result of this construction, one finds the cohomology group $H^2(X_{Schoen}, \ZZ) \simeq \text{Pic}(X_{Schoen}) $ to be 
\begin{align}
    H^2(X_{Schoen}, \ZZ) = \frac{H^2(S^1, \ZZ)\oplus H^2(S_2, \ZZ)}{H^2(\PP^1, \ZZ)}\, .
\end{align}
In terms of the resolved orbifold in the present convention, the base of fibre product is torus one. Along this direction the two factors of the product are glued together such that 
\begin{align}
    x_1\in S_1: \; x_1=(z_1, z_3)\, , && x_2\in S_2: \; x_2=(z_1, z_2)\, ,
\end{align}
and $\pi_i(x_i)=z_1$, where $z_i$, $i=1, 2, 3$, are the complexified coordinates of the three tori, which are subject to the orbifolding group identifications.

In the following we discuss the different F-theory lifts of this geometry. As it is symmetric in torus $\TT^2_2$ and $\TT^2_3$ or equivalently $S_1$ and $S_2$, there are two inequivalent lifts. 

\paragraph{F-theory lift along $\TT^2_1$} In this lift, the shift only acts in the F-theory fibre, that is the common $\PP^1$ is lifted. 

As the base of the F-theory fibration is not effected by the shift, it remains almost the same as in the previous example $\ZZ_1\times\ZZ_1-1-1$ without shift. In both fixed point sectors ($\omega$ and $\vartheta\omega$) there are four fixed tori, that mutually intersect.

The shift in the fibre identifies fixed points and thus resolution divisors, compared to $\ZZ_2\times\ZZ_2-1-1$. Former $D_4$-type fibres are folded into $B_2^{(2)}\simeq C_2^{(2)}$-type. This folding is a twisted affine folding, as explained in Section~\ref{sec:new-features}, and yields the gauge group supported by the fixed point divisors are $SO(5)\simeq Sp(2)$. Note that the singularity at the intersection of these divisors does not exceed the minimality bound and yields therefore no conformal matter.

Beyond this, there are fixed points, when $\vartheta$ is restricted to the base tori. As explained above in Section~\ref{sec:new-features}, this yields SCFT sectors in the theory with $\Nn=(2, 0)$ SUSY, which also contribute to the massless spectrum. In our analysis of the geometry \textit{the number} of these sectors, however, does not match with what anomaly cancellation requires. The restricted element $\vartheta|_\text{base}$ acts as 
\begin{align}
    \vartheta|_\text{base}:\;\;\; \left(\begin{array}{cc|cccc}
        1/2 & -1/2 \;\;&\; 0 & 0 & 0 & 0
    \end{array}\right)\, ,
\end{align}
which has $4\times4=16$ fixed points. Requiring the cancellation of the irreducible gravitational anomaly with the data provided below, restricts the number of $\Aa_1$ factors to $8$. This mismatch is likely due to an additional geometric effect that reduces the number of $\Nn=(2, 0)$ SCFT sectors that are found from the fixed points of $\vartheta|_\text{base}$ either by an identification of divisors in the base or an effect similar to that encountered in the analysis of the lift of torus $\TT^2_3$ of the orbifold $\ZZ_3\times\ZZ_3-1-2$. This issue can be analyzed in detail in the resolved phase of the geometry, which goes beyond the scope of this work.

Concluding, the massless sector of the theory is summarized as
\begin{align}
    \begin{array}{|c|c|} 
    \hline 
    G & SO(5)^4\times SO(5)^4 \\ \hline
    H  & (4+4)\times \mathbf{10} +19+1 \\ \hline
    V & (4+4)\times \mathbf{10}   \\ \hline
    T & 1  \\  \hline
    T_{2, 0} & 8\times \Aa_1  \\  \hline
    \end{array} \, .\nonumber
\end{align}
Note, there is charged matter, that contributes to the number of hypers, which is in the adjoint representation of $SO(5)$. The appearance of these multiplets is consistent as the base divisors associated to the fixed points of $\omega$ and $\vartheta\omega$ are genus-one curves. Cancellation of the gauge anomaly of $SO(5)$ on such curves requires exactly one adjoint multiplet. Finally we also note, that the identification of the rank $8\times 2$ gauge group and its single tensor is consistent with the $19$ K\"ahler parameters. Note again that the $\mathcal{A}_1$ theories do not contribute to those, but are again a hallmark for a $\ZZ_2$ discrete 0-form gauge symmetry. 

\paragraph{F-theory lift along $\TT^2_2$} Lifting either torus two or three yields an F-theory fibration, where the shift acts in one of the two base tori. In terms of the fibre product (equation~\eqref{equ:Schoen-prod}), the fibre of one of the two factors chosen to be F-theory fibre. Note, that this lift has already been discussed in \cite{Gopakumar1996}. It is nevertheless briefly discussed for the sake of completeness.

The fixed point structure yields two genus-one curves in the base over which the fibre is of $D_4$ type. This exhausts the amount of co-dimension one singularities in the base. Comparing to the $\ZZ_2\times\ZZ_2$ orbifold with trivial affine class, one realizes that the amount of fixed tori in the $\omega$-sector is reduced by a pairwise identification of fixed points in the base. Furthermore, the fixed tori due to the action of $\vartheta$, that mutually intersected the former, do not appear here, as $\vartheta$ acts freely on the orbifold as well as on the base.

Note, that this analysis is consistent with the fact, that the resolved orbifold is a Schoen manifold. There we have two $dP_9$ surfaces with homology lattice $\begin{psmallmatrix}  0&1\\1&0\end{psmallmatrix}\oplus E_8$. These are glued along the common $\PP^1$ whose homology class is encoded in the transcendental $U$ lattice part, as described in the beginning of this section. This suggests, that in the F-theory compactification that follows the fibration structure of the fibre product, one finds a gauge group which is either $E_8$ or a rank preserving subgroup thereof. Consistently, the (adjoint) higgsing of $E_8$ to $SO(8)\times SO(8)$ yields two $\mathbf{28}$ adjoint multiplets matching with gauge anomaly cancellation of $SO(8)$ on a genus-one curve. To summarize we have the massless spectrum
\begin{align}
    \begin{array}{|c|c|} 
    \hline 
    G & SO(8)^2 \\ \hline
    H  & 2\times \mathbf{28} +19+1 \\ \hline
    V & 2\times \mathbf{28}   \\ \hline
    T & 9  \\  \hline
    \end{array} \, .\nonumber
\end{align}
Note that the nine tensor multiplets are geometrically realized by the universal mode as well as the volumes of the resolution divisors of the singularities associated to the eight fixed points of the action of $\omega$. Again, the rank $2 \times 4$ gauge group, including the $9$ tensors is consistent with the 19 K\"ahler parameters found in the geometry.

\subsection*{$\ZZ_2\times\ZZ_2-1-3$ orbifold}
Turning to the second non-trivial affine class, two shifts are introduced into the generator $\vartheta$ yielding the three inequivalent orbifolding group elements
\begin{align}
    \begin{array}{c}
        \vartheta:\\\omega:\\\vartheta\omega:
    \end{array}
    \;\;\;
	\begin{array}{ccc|ccccccc}
		\left(\right.\;\;0\;\;\; & 1/2 & -1/2 \;\;&\; 0 & 1/2 & \;0 & 0\; & 0 & 1/2 \left.\right)  \\
		\left(\right.1/2 & 0 & -1/2 \;\;&\; 0 & 0 & \;0 & 0\; & 0 & \;\;0\; \left.\right) \\
		\left(\right.1/2 & 1/2 & \;0 &\; 0 & 1/2 & \;0 & 0\; & 0 & 1/2 \left.\right) 
	\end{array}\, .
\end{align}
The resulting geometry has Hodge numbers 
\begin{align}
	(h^{(1, 1)}_{\text{untw.}}, h^{(1, 1)}_{\text{tw.}}) = (3, 8)\, , && 	(h^{(2, 1)}_{\text{untw.}}, h^{(2, 1)}_{\text{tw.}}) = (3, 8)\, .
\end{align}

\begin{figure}[t]
	\centering
	\includegraphics[width=0.6\linewidth]{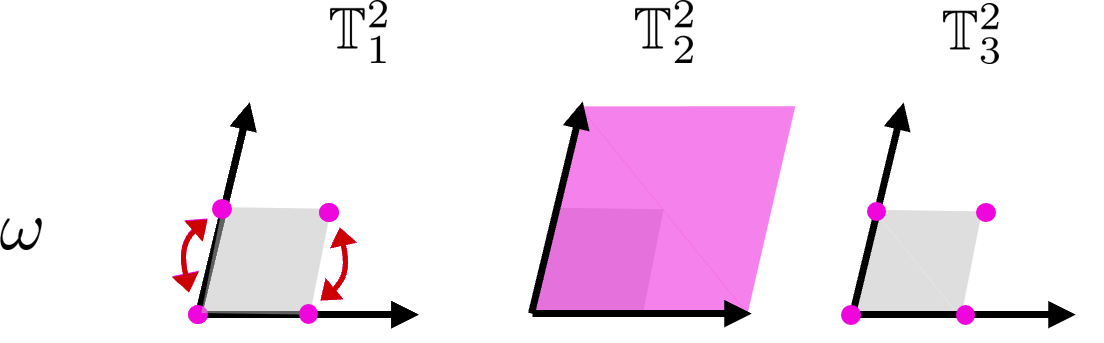}
	\caption{An illustration of the fixed points (pink) in the $\ZZ_2\times\ZZ_2-1-3$ orbifold.}
	\label{fig:Z2xZ2-1-3_FPinTori_2}
\end{figure}

Note that the action of the orbifolding group is symmetric in torus one and three. This suggests that there are only two inequivalent lifts, $\TT^2_1$ (or $\TT^2_3$) and $\TT^2_2$. This is, naively, also reflected in the fixed point structure (see Figure~\ref{fig:Z2xZ2-1-3_FPinTori_2}). However, in the identification of fixed points due to the shifts there appears to be a choice. The only remaining fixed points are due to the action of $\omega$, as the action of the other orbifolding group elements on the orbifold is smooth. The action naively has $4\times4=16$ fixed points. However, conjugation yields that there are in fact only $8$.\footnote{This agrees with $h^{(1, 1)}$ obtained \cite{Fischer_2013} or through data of the orbifolder \cite{Nilles_2012}.} This identification by the action of the shift inevitably causes a asymmetry between torus one and three, as the identification only acts in one of these tori. As shown below, depending on the choice of identification and lift, we find theories consistent with the cancellation of the irreducible gravitational anomaly with either two $SO(8)$ or four $SO(5)=Sp(2)$. 

In the following all possibilities of choice of F-theory fibre are discussed, supposing the fixed point structure illustrated in Figure~\ref{fig:Z2xZ2-1-3_FPinTori_2}.
\paragraph{F-theory lift along $\TT^2_2$} In this lift, the shift acts in both base tori and there is no ambiguity. Descending from the $\TT^2_2$ lift of the $\ZZ_2\times\ZZ_2-1-2$ orbifold, this model is also discussed in \cite{Gopakumar1996}.

As both $\vartheta$ and $\vartheta\omega$ act freely on the orbifold, there are no fixed points of co-dimension one in the base. In fact, $G=\emptyset$ and, furthermore, 
\begin{align}
    \begin{array}{|c|c|} 
    \hline 
    G & \emptyset \\ \hline
    H  & 11+1 \\ \hline
    V & -   \\ \hline
    T & 9  \\  \hline
    \end{array} \, .\nonumber
\end{align}
As in the lift of $\TT^2_2$ of the $\ZZ_2\times\ZZ_2-1-2$ orbifold, the tensors are due to the eight fixed points of the action of $\omega$ and the universal mode. Also the 11 K\"ahler moduli are identified with the nine tensors and generic fiber and base class.

\paragraph{F-theory lift along $\TT^2_3$} Lifting torus three to be F-theory fibre and identifying fixed points in the base yields two fixed tori. The degeneration of the fibre suggests, that one finds $SO(8)$ gauge group factors here. These are actually genus-one curves, as one observes that there are no fixed points of co-dimension two in the base, that intersect the fixed tori. To anticipate the analysis of the gauge anomaly on these curves, one finds two adjoint charged hyper multiplets of the $SO(8)$ in the spectrum.

Although this exhausts the fixed points of the total space, we again encounter some fixed point elements coming from freely acting elements of the space group
$\vartheta\omega$, which, as mentioned above, in fact do not intersect the fixed tori. As described in Section~\ref{sec:new-features}, over these locations in the base, one observes a multiple fibre giving rise to $\Nn=(2, 0)$ SCFT sectors of the theory. As there are $4\times 4/2$ such loci due to identification, these $\Aa_1$ sectors yield further multiplets, equivalent to $8\times 30=240$ hypers. 

The remaining orbifolding group element acts freely on the orbifold as well as the base. 

Counting multiplets, one finds cancellation of the irreducible gravitational anomaly as well as gauge anomalies, as 
\begin{align}
    \begin{array}{|c|c|} 
    \hline 
    G & SO(8)^2 \\ \hline
    H  & 2\times \mathbf{28} +11+1 \\ \hline
    V & 2\times \mathbf{28}   \\ \hline
    T & 1  \\  \hline
    T_{2, 0} & 8\times \Aa_1  \\  \hline
    \end{array} \, .\nonumber
\end{align}
In order to show cancellation of the reducible gravitational anomaly the entire intersection pairing is required. This, however, still remains an open task due issues connected to the unimodularity of the embedding of $\Aa_1$ into the theory.

\paragraph{F-theory lift along $\TT^2_2$} In this case, the identification of fixed points takes place in the F-theory fibre, which gives rise to twisted affine foldings of Dynkin diagrams in the F-theory interpretation of the geometry. 

Investigating the F-theory base, one finds four inequivalent fixed tori associated to the orbifolding group element $\omega$. These are, further, not intersected by any singular structure in the geometry, yielding a genus-one curve in the base. The fibre here is a folding of the affine Dynkin diagram $D_4$ to a twisted affine Dynkin diagram $B_2^{(2)}\simeq C_2^{(2)}$, yielding an $SO(5)\simeq Sp(2)$ gauge group contribution. Another example of a twisted affine folding induced by a roto-translation. Beyond that analogously to the lift discussed above, eight $\Nn=(2, 0)$ SCFT sectors of type $\Aa_1$ are found. 

To conclude, the massless sprectrum of this theory given by
\begin{align}
    \begin{array}{|c|c|} 
    \hline 
    G & SO(5)^4 \\ \hline
    H  & 4\times \mathbf{10} +11+1 \\ \hline
    V & 4\times \mathbf{10}   \\ \hline
    T & 1  \\  \hline
    T_{2, 0} & 8\times \Aa_1  \\  \hline
    \end{array} \, .\nonumber
\end{align}
This spectrum cancels the irreducible gravitational anomaly as well as gauge anomalies. The verification of the cancellation of the reducible gravitational anomaly again is left for future studies.

\subsection*{$\ZZ_2\times\ZZ_2-1-4$ orbifold}
Analyzing the maximal affine class of $\ZZ_2\times \ZZ_2$ orbifolds on a lattice $\Lambda= SU(2)^6$, one introduces a third shift by setting 
\begin{align}
    \begin{array}{c}
        \vartheta:\\\omega:\\\vartheta\omega:
    \end{array}
    \;\;\;
	\begin{array}{ccc|ccccccc}
		\left(\right.\;\;0\;\;\; & 1/2 & -1/2 \;\;&\; 0 & 1/2 & \;0 & 0\; & 0 & 1/2 \left.\right)  \\
		\left(\right.1/2 & 0 & -1/2 \;\;&\; 0 & 0 & \;0 & 1/2 & 0 & \;\;0\; \left.\right) \\
		\left(\right.1/2 & 1/2 & \;0 &\; 0 & 1/2 & \;0 & 1/2 & 0 & 1/2 \left.\right) 
	\end{array}\, ,
\end{align}
yielding a geometry with Hodge numbers
\begin{align}
	(h^{(1, 1)}_{\text{untw.}}, h^{(1, 1)}_{\text{tw.}}) = (3, 0)\, , && 	(h^{(2, 1)}_{\text{untw.}}, h^{(2, 1)}_{\text{tw.}}) = (3, 0)\, .
\end{align}
The action of all those generators on the total space orbifold is smooth. Irrespective of the chosen lift the only structure remaining here are eight fixed points of the restriction of one of the orbifolding group elements to the base. As in the above examples, this yields $8\times \Aa_1$ $(2,0)$ subsector, consistent with the cancellation of the irreducible gravitational anomaly via the following massless spectrum
\begin{align}
    \begin{array}{|c|c|} 
    \hline 
    G & \emptyset \\ \hline
    H  & 3+1 \\ \hline
    V & -   \\ \hline
    T & 1  \\  \hline
    T_{2, 0} & 8\times \Aa_1  \\  \hline
    \end{array} \, .\nonumber
\end{align}

\newpage

\bibliographystyle{JHEP}
\bibliography{main}

\end{document}